\documentclass[useAMS,usenatbib]{mn2e}

\usepackage{natbib}
\usepackage{cite}
\usepackage{amssymb}
\usepackage{graphicx}
\usepackage{graphics}
\usepackage{lscape}
\usepackage{color}
\usepackage{verbatim}

\newcommand{\ha}{H$\alpha$}

\newcommand{\zsun}{$Z_\odot$}
\newcommand{\msun}{$M_\odot$}

\newcommand{\hi}{H\,{\sc i}}
\newcommand{\hii}{H\,{\sc ii}}

\newcommand{\nii}{[N\,{\sc ii}]}

\newcommand{\oiii}{[O\,{\sc iii}]}
\newcommand{\oii}{[O\,{\sc ii}]}

\newcommand{\sii}{[S\,{\sc ii}]}
\newcommand{\ariii}{[Ar\,{\sc iii}]}

\newcommand{\te}{T$_e$}

\newcommand{\hbeta}{H$\beta$}
\newcommand{\halpha}{H$\alpha$}

\newcommand{\lin}{$\,\lambda$}
\newcommand{\llin}{$\,\lambda\lambda$}

\newcommand{\oh}{12\,+\,log(O/H)}

\newcommand{\ohsun}{\mbox{12\,+\,log(O/H)$_\odot\,=\,$}}

\newcommand{\vs}{vs.}

\title[H\,{\large II} Regions in the XUV disc of NGC~4625]{On the Nature of the H\,{\LARGE II} Regions in the Extended Ultraviolet Disc of NGC~4625}
\author[Q.E. Goddard et al.]{Q.E. Goddard$^{1}$, F. Bresolin$^{2}$\thanks{E-mail: bresolin@ifa.hawaii.edu}, R.C. Kennicutt$^{1}$, E.V. Ryan-Weber$^{3}$ \& 
\newauthor F.F. Rosales-Ortega$^{1}$ \\
$^{1}$Institute of Astronomy, University of Cambridge, Madingley Road, Cambridge CB3 0HA, UK\\
$^{2}$Institute for Astronomy, 2680 Woodlawn Drive, Honolulu, HI 96822, USA \\
$^{3}$Centre for Astrophysics \& Supercomputing, Swinburne University of Technology, Mail H39, PO Box 218, Hawthorn, 3122 VIC, Australia}

\begin{document}

\date{}

\pagerange{\pageref{firstpage}--\pageref{lastpage}} \pubyear{2010}

\maketitle

\label{firstpage}

\begin{abstract}
Using deep Subaru/FOCAS spectra of 34 \hii\ regions in both the inner  and outer parts of the extended ultraviolet (XUV) disc galaxy NGC~4625 we have measured an abundance gradient out to almost 2.5 times the optical isophotal radius. We applied several strong line abundance calibrations to determine the \hii\ region abundances, including R$_{23}$, \nii/\oii, \nii/\ha\ as well as the \oiii\lin4363 auroral line, which we detected in three of the \hii\ regions. We find that at the transition between the inner and outer disc the abundance gradient becomes flatter. In addition, there appears to be an abundance discontinuity  in proximity of this transition. Several of our target \hii\ regions appear to deviate from the ionisation sequence defined in the \nii/\ha\ vs. \oiii/H$\beta$ diagnostic diagram by bright extragalactic \hii\ regions. Using theoretical models we conclude that the most likely explanations for these deviations are either related to the time evolution of the \hii\ regions, or stochastic variations in the ionising stellar populations of these low mass \hii\ regions, although we are unable to distinguish between these two effects. Such effects can also impact on the reliability of the strong line abundance determinations.

\end{abstract}


\section{Introduction}

Extragalactic \hii\ regions act like galactic buoys, highlighting sites of recent or ongoing star formation. It has long been known that \hii\ regions are not just found in the active star forming discs of galaxies, but also at large galactocentric radii, as shown, for example, by \citet{ferguson98b} and \citet{lelievre00}. These observations have been bolstered by the discovery of young B-type stars at large galactocentric radii \citep{cuillandre01,davidge07}. However, the idea of star formation occurring beyond the optical edge of a galaxy (commonly defined by the 25th magnitude B-band isophotal radius, R$_{25}$) was revolutionised with the advent of the \emph{Galaxy Evolution Explorer} (GALEX) satellite \citep{martin03}. The use of GALEX as part of the Nearby Galaxies Survey \citep{bianchi03} revealed numerous galaxies where the UV emission extends well beyond the optical edge. These galaxies have since been termed `extended ultraviolet disc' (XUV disc) galaxies \citep{thilker05, thilker07a,gildepaz05}. 

The GALEX satellite, in conjunction with optical and infrared observations, has given us some major insights into the nature of these XUV discs. We now know that these structures are rather common, with an incident rate $\gtrsim 30\%$ in spiral galaxies \citep{zaritsky07,thilker07a}. XUV discs are complex and varied systems, divided by \citet{thilker07a}  into two types. Type 1 XUV discs are highly structured, displaying  UV-bright, yet optically dim, knots. Type 2 XUV discs, which are predominantly found in late-type spirals, show an extended UV-bright, but once again optically faint, region beyond the galaxy edge. Many of the UV knots correspond to structures in the \hi\ distribution \citep{gildepaz05,thilker05}, and some  have \hii\ region counterparts \citep{gildepaz05,goddard10}. The UV and \ha\ fluxes of these \hii\ regions indicate a mass of the underlying stellar population between $10^{3}$ and 10$^{4}$~\msun. These studies also show that ionisation is dominated by only a handful or a single O-type star \citep{gildepaz05,bresolin09,goddard10}. The ages of the UV complexes in XUV discs appear to range between a few Myr and several hundred Myr \citep{zaritsky07,dong08}.

Even though the mean surface gas density at extreme galactocentric radii falls below the critical threshold for star formation  \citep{kennicutt89}, local  density perturbations may still promote star forming activity in the outer discs of spiral  galaxies \citep{martin01}.
This is supported by models of the propagation of spiral density waves into the extended discs \citep{bush08}. However, other mechanisms may also play a part. \citet{gildepaz05}, for example, suggested that an interaction between NGC~4625 and NGC~4618 acted as a trigger for star formation at extreme radii in NGC~4625. \citet{elmegreen06} proposed a range of other mechanisms, including stellar compression and turbulence compression among others.

\hii\ regions in XUV discs are ideal tracers of star-formation at extreme radii, but they also encode information about the kinematics of these structures \citep{christlein08} and the chemical evolution of these extreme star formation environments \citep{gildepaz07b,bresolin09}. Previous studies of the abundances of XUV disc \hii\ regions have insofar been rather limited. \citet{ferguson98a} studied the galaxies NGC~628, NGC~1058 and NGC~6946,  but their sample was limited to nine \hii\ regions beyond R$_{25}$. More in depth studies of \hii\ region abundances in XUV discs have focused on two galaxies in particular. \citet[hereafter referred to as G07]{gildepaz07b}  studied 31 \hii\ regions in the XUV discs of M83 and NGC~4625, finding in both cases an exponential abundance gradient extending into the outer disc. However, a deeper study of M83 by \citet{bresolin09}, containing 49 \hii\ regions out to 2.6 R$_{25}$, found a marked discontinuity in the abundance gradient at  the optical edge of the galaxy, together with a flat gradient in the XUV disc.

NGC~4625 and M83 have been the most studied XUV disc galaxies, partly due to their close proximity. These galaxies possess some of the most XUV discs known, extending out to 4~R$_{25}$ in both cases \citep{thilker05,gildepaz05}. They share similar trends in their \ha\ and UV radial profiles, with a turnover in the \ha\ profile near R$_{25}$, yet a smooth far-UV profile extending into the XUV disc \citep{goddard10}. However,  M83 displays a large number of  UV knots in the XUV disc in association with structured \hi\ emission, whilst the  lower luminosity NGC~4625 has a more diffuse UV emission accompanying a low surface brightness optical component \citep{thilker07a}. 

In this paper we present a detailed chemical abundance study of the \hii\ regions in NGC~4625, including  targets across the XUV disc
not previously  observed by G07. 
We have sampled \hii\ regions in both the inner and outer disc out to $\sim2.5$ R$_{25}$, and study their  oxygen abundances  based on several strong nebular emission line indicators. 
We first describe our observations and the data reduction in \S \ref{sec:obs}. In \S \ref{sec:nebabunds} we discuss the oxygen abundances that we obtained and compare our results to those of previous datasets. \S \ref{sec:roguehii} discusses the nature of \hii\ regions in the XUV disc, considering the effects of an ageing population and stochasticity on the observed nebular line ratios. We briefly discuss our findings in the context of extragalactic star formation  in \S \ref{sec:interp}, before we finally summarise our conclusions. Throughout this paper we assume a distance to NGC~4625 of 9.5~Mpc  \citep{kennicutt03}, an inclination angle $i=27^{\circ}$ and a position angle of the major axis $\theta=150^{\circ}$ \citep{bush04}. The R$_{25}$ value of 66~arcsec, from \citet{devaucouleurs91}, corresponds to 3.04~kpc. We adopt a solar metallicity value  \ohsun 8.69 \citep{asplund09}.

\section{Observations and Data Reduction} \label{sec:obs}

\begin{table}
 \caption{\hii\ region sample.}
 \label{tab:posinfo}
 \begin{center}
 \begin{tabular}{@{}l c c c}
  \hline
  \hline
  \phantom{aaaaa}ID\phantom{aaa} & \phantom{aaaa}R.A.\phantom{aaaa} & \phantom{aaaa}Decl. \phantom{aaaa} & R/R$_{25}$ \\
   & (J2000.0) & (J2000.0) & \\
 \phantom{aaaaa}(1) & (2) & (3) & (4) \\
  \hline
 103 \hfill({\sc xuv}-12) & 12:41:42.93 & 41:16:37.8 &  1.89 \\
104.204 & 12:41:44.43 & 41:17:15.3 &  1.70 \\
105 & 12:41:45.24 & 41:17:34.5 &  1.70 \\
106 & 12:41:45.94 & 41:15:58.2 &  1.44 \\
107.207 & 12:41:47.45 & 41:15:28.8 &  1.46 \\
108 \hfill({\sc xuv}-10) & 12:41:48.64 & 41:17:46.6 &  1.43 \\
110a & 12:41:50.30 & 41:16:10.0 &  0.57 \\
113 & 12:41:51.34 & 41:16:36.0 &  0.29 \\
114 & 12:41:51.91 & 41:16:37.4 &  0.22 \\
115 & 12:41:52.70 & 41:16:23.6 &  0.04 \\
116a & 12:41:53.02 & 41:16:33.9 &  0.14 \\
116b & 12:41:53.13 & 41:16:33.9 &  0.15 \\
116c & 12:41:53.36 & 41:16:34.5 &  0.19 \\
117 \hfill({\sc xuv}-8) & 12:41:54.28 & 41:17:26.8 &  1.05 \\
118 \hfill({\sc xuv}-7) & 12:41:56.21 & 41:15:30.9 &  1.05 \\
121 & 12:41:58.06 & 41:15:43.8 &  1.17 \\
122 \hfill({\sc xuv}-5) & 12:41:59.14 & 41:15:21.3 &  1.52 \\
123 \hfill({\sc xuv}-4) & 12:42:00.33 & 41:16:39.7 &  1.52 \\
124 & 12:42:01.38 & 41:17:11.6 &  1.92 \\
128.225 \hfill({\sc xuv}-3) & 12:42:04.04 & 41:16:35.7 &  2.22 \\
130 \hfill({\sc xuv}-2) & 12:42:05.11 & 41:16:12.1 &  2.39 \\
131.228 & 12:42:06.40 & 41:16:55.6 &  2.76 \\
203 & 12:41:43.11 & 41:16:48.0 &  1.86 \\
205 & 12:41:45.29 & 41:17:17.2 &  1.57 \\
208 & 12:41:49.00 & 41:15:22.0 &  1.33 \\
210a & 12:41:50.29 & 41:16:18.7 &  0.50 \\
213 & 12:41:51.50 & 41:16:27.3 &  0.23 \\
214 & 12:41:52.06 & 41:16:21.1 &  0.16 \\
215 & 12:41:52.65 & 41:16:06.9 &  0.31 \\
216 & 12:41:53.80 & 41:17:02.0 &  0.63 \\
217 & 12:41:54.64 & 41:16:12.1 &  0.41 \\
221a & 12:42:00.06 & 41:15:23.1 &  1.64 \\
221b & 12:41:59.49 & 41:15:23.3 &  1.55 \\
223 & 12:42:02.02 & 41:16:19.1 &  1.80 \\
\hline
\end{tabular}
\end{center}
 \medskip (1) Identification number; a compound number (e.g. 131.228) indicates an \hii\ region observed with both our multi object  masks; ID numbers with a suffix (e.g.~116c) indicate that more than one \hii\ region was extracted from the same slit. {\sc xuv} numbers identify the objects studied by G07.
 (2, 3) Right Ascension and Declination (J2000). (4) Galactocentric distance in units of the isophotal radius R$_{25}$ (R$_{25}$~=~66\arcsec~=~3.04~kpc).
\end{table}

Our \hii\ region selection was based on deep \ha\ images taken using the Faint Object Camera And Spectrograph (FOCAS, \citealt{kashikawa}) at the Subaru telescope situated on Mauna Kea in Hawaii. The field of view of the instrument (6 arcmin) encompasses the entire XUV disc of NGC~4625.
A continuum-subtracted \ha\ image was constructed using a 900\,s on-band image in combination with a 120\,s continuum image. We 
designed two multi-object slit masks, attempting  to cover as wide a range in galactocentric radius as possible, and including several \hii\ regions which were part of the spectroscopic study by G07, to enable a comparison of identical \hii\ regions.

The spectra were obtained with FOCAS on the night of March 22, 2009, using 1\arcsec\ (= 46 pc) slits in combination with the 300R grism (used in second order) to cover the blue part of the spectrum (approximately from 3400\,\AA\ to 5300\,\AA) and the VPH650 grism for the red part of the spectrum (5300\,\AA\ to 7700\,\AA). The spectral resolution, estimated from the gaussian FWHM of lines in the arc lamp frames, is approximately 5.5\,\AA. We exposed for a total time of 2 hours in the blue and 80 minutes in the red for each mask. Seeing conditions varied through the night from a poor 2\arcsec\ to a more favourable 0\farcs8 on occasion.  The variations in the atmospheric conditions occurred mostly during the observations taken with the first mask, while
the seeing stabilized around 1\arcsec\ for the  remaining part of the night.
To flux-calibrate our spectra we took long slit spectra of four standard stars, GD71 and G191-B2B at evening twilight, and HZ44  and BD+33d2642 at the end of the night. 
The data were reduced using standard {\sc iraf\footnotemark} routines. 
\footnotetext{IRAF is distributed by the National Optical Astronomy Observatory, which is operated by the Association of Universities for Research in Astronomy, Inc., under co-operative agreement with the National Science Foundation.}

\begin{figure*}
   \centering
   \includegraphics[width = 8.8cm]{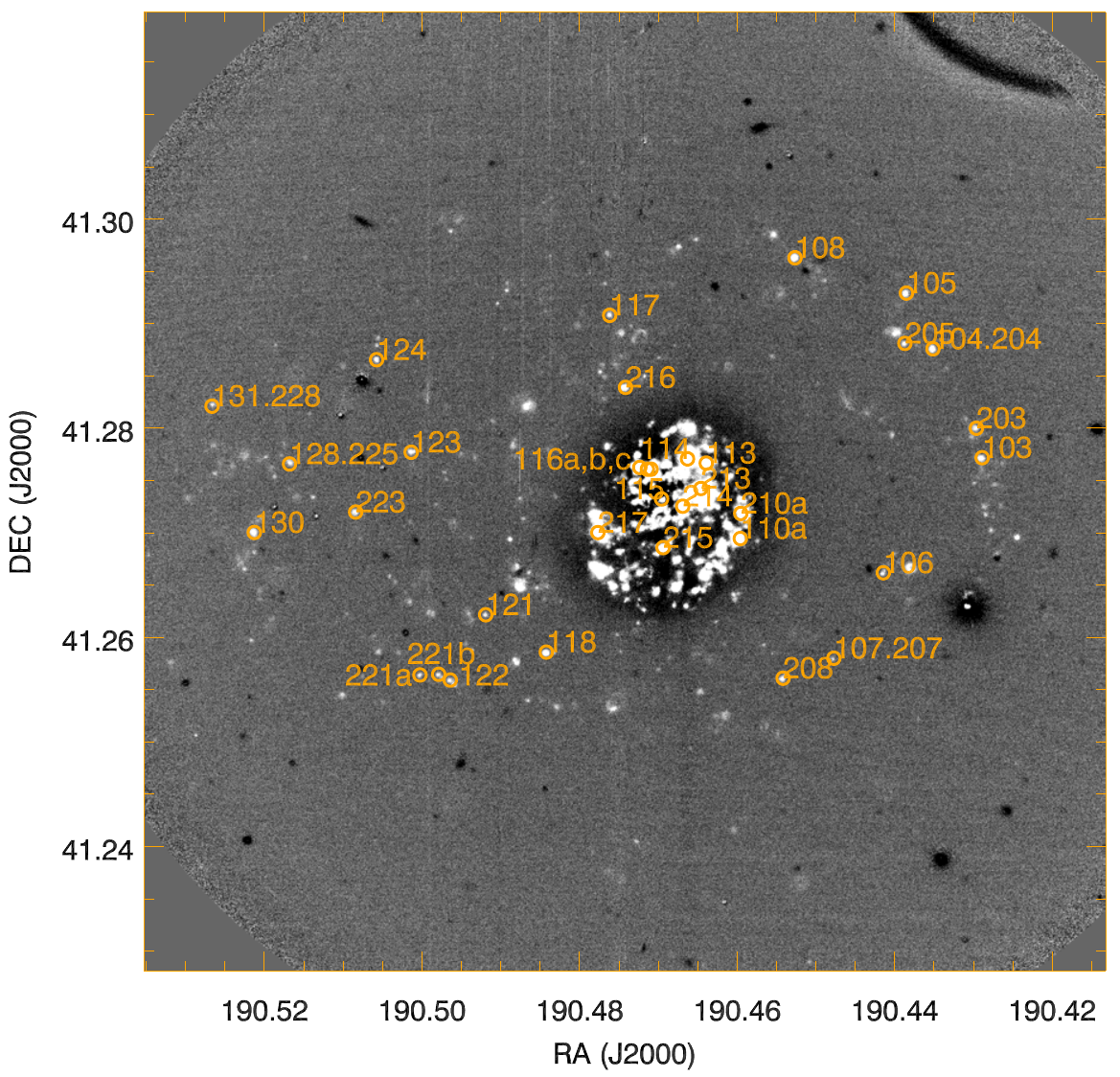}
   \includegraphics[width = 8.8cm]{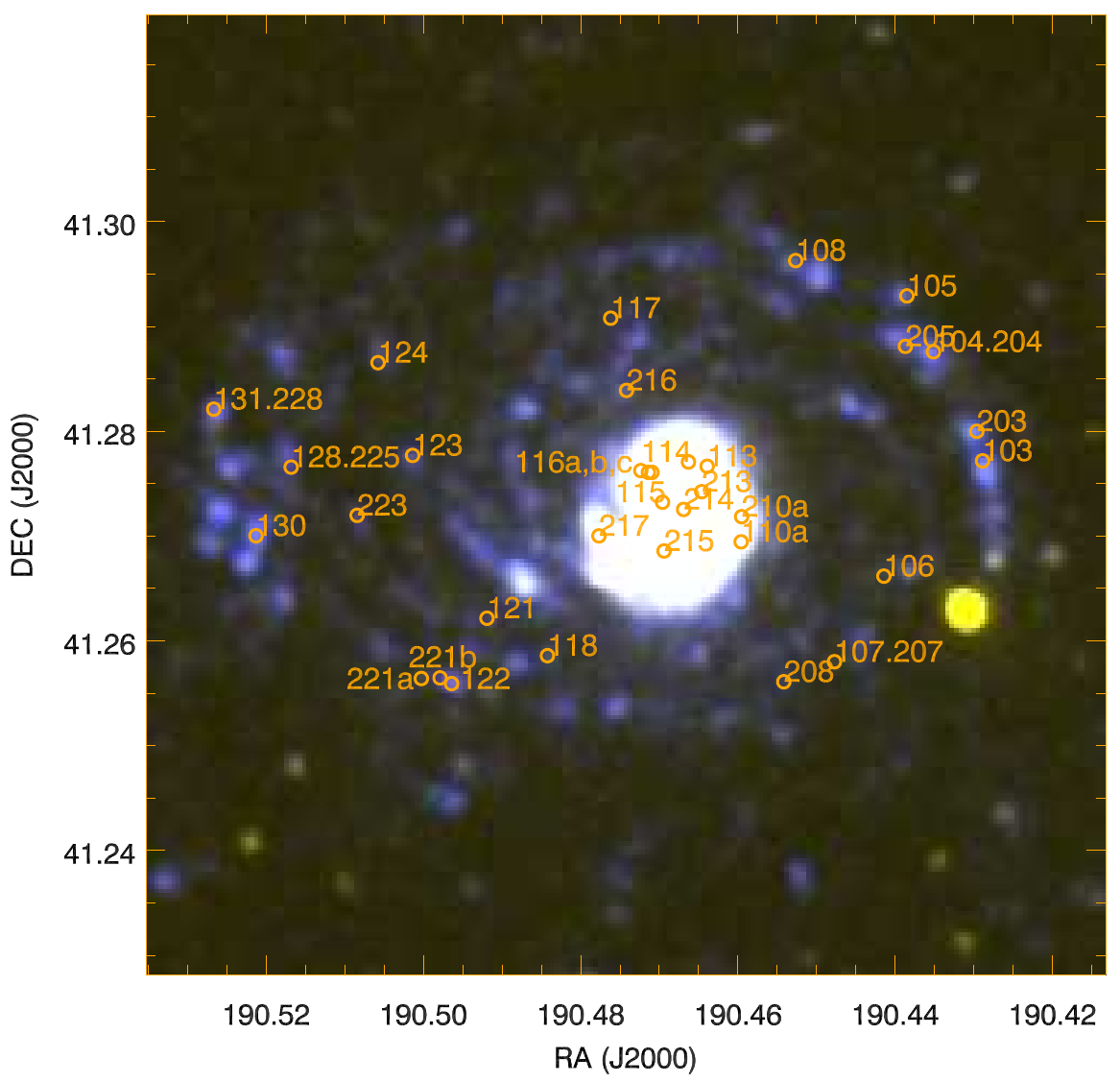} 
   \caption{{\em Left:} Continuum-subtracted \ha\ image of NGC~4625, obtained from our Subaru FOCAS imaging. {\em Right:} GALEX UV colour composite image of NGC~4625. Our \hii\ region sample  is shown by the orange circles marked with the object IDs.}
   \label{fig:map}
\end{figure*}

We obtained spectra for 34 \hii\ regions from the two slit masks combined. We list their positions and galactocentric distances
in Table~\ref{tab:posinfo}. Our targets are marked  on a continuum-subtracted \ha\ image and a
GALEX ultraviolet colour composite image of NGC~4625 in Fig.~\ref{fig:map}. Four of the \hii\ regions were observed in both masks. 
A few of the slits, identified by a letter following the slit identification number in Table~\ref{tab:posinfo}, included multiple targets.

We measured the spectral line strengths  by direct integration of the flux measured between two continuum regions interactively selected on each side of the line.
The reddening correction was done using the extinction curve published by \citet{osterbrock}, and assuming an intrinsic H$\gamma$/H$\beta$ ratio of 0.47, which holds at \te~=~$10^{4}$~K. We were unable to use the H$\alpha$/H$\beta$ ratio because these two Balmer lines appear separately in the red and blue spectra, respectively. Therefore, the theoretical ratio H$\alpha$/H$\beta$ $= 2.86$  was adopted to renormalise the line fluxes in the red spectral range, after we had made the reddening correction.
The line flux ratios that involve emission features observed separately in the blue and red spectra could potentially be affected by the variations in the seeing conditions mentioned earlier. Fortunately, the spectra for our second mask were obtained with an approximately constant image quality ($\sim 1"$). Moreover, since a few of the targets were included in both masks, we could verify that the 
flux ratios did not suffer from changes in the seeing.

\begin{table*}

 \caption{Reddening-corrected lines fluxes.}
 \label{tab:galres}
 \begin{center}
 \resizebox{17.5cm}{!} { 
 \begin{tabular}{@{}l c c c c c c c c c}
  \hline
  \hline
ID & \oii & \oiii & \nii & \sii & \ariii & F(H$\beta$) / $10^{-16}$ & c(H$\beta$) & $\log\frac{\mbox{\nii}}{\mbox{\oii}}$ & $\log\frac{\mbox{\nii}}{\mbox{H}\alpha}$ \\
 & 3727 & 5007 & 6583 & 6717+6731 & 7135 & (erg\,s$^{-1}$\,cm$^{-2}$) & & & \\
 (1) & (2) & (3) & (4) & (5) & (6) & (7) & (8) & (9) & (10) \\
  \hline
       103 & $  311 \pm    21$ & $   92 \pm     4$ & $   29 \pm     1$ & $   67 \pm     2$ & $   4 \pm 0.9$ &     3.2 & $  0.00 \pm   0.20$ & $ -1.02$ & $ -0.98$ \\
   104.204 & $  344 \pm    16$ & $  122 \pm     5$ & $   28 \pm     1$ & $   43 \pm     1$ & $   6 \pm 0.6$ &     3.9 & $  0.00 \pm   0.14$ & $ -1.08$ & $ -1.00$ \\
       105 & $  295 \pm    25$ & $  158 \pm     7$ & $   24 \pm     1$ & $   37 \pm     2$ & $   7 \pm 1.3$ &     2.4 & $  0.00 \pm   0.20$ & $ -1.08$ & $ -1.07$ \\
       106 & $  291 \pm    38$ & $  337 \pm    19$ & $   26 \pm     3$ & $   49 \pm     5$ & $  18 \pm 3.6$ &     1.0 & $  0.00 \pm   0.29$ & $ -1.05$ & $ -1.04$ \\
   107.207 & $  333 \pm    40$ & $  158 \pm    10$ & $   41 \pm     4$ & $   61 \pm     5$ & ... &     0.6 & $  0.00 \pm   0.29$ & $ -0.90$ & $ -0.84$ \\
       108 & $  266 \pm    16$ & $  414 \pm    19$ & $   26 \pm     1$ & $   68 \pm     2$ & $   9 \pm 1.0$ &     4.4 & $  0.00 \pm   0.17$ & $ -1.01$ & $ -1.04$ \\
      110a & $  148 \pm    25$ & $   20 \pm     1$ & $   98 \pm     2$ & $   78 \pm     1$ & $   4 \pm 0.1$ &    17.2 & $  0.16 \pm   0.16$ & $ -0.18$ & $ -0.46$ \\
       113 & $  139 \pm    21$ & $   25 \pm     1$ & $   96 \pm     2$ & $   40 \pm     1$ & $   7 \pm 0.2$ &    17.7 & $  0.96 \pm   0.18$ & $ -0.16$ & $ -0.47$ \\
       114 & $  155 \pm    12$ & $   24 \pm     1$ & $   97 \pm     1$ & $   67 \pm     1$ & $   4 \pm 0.1$ &    53.8 & $  0.08 \pm   0.13$ & $ -0.20$ & $ -0.47$ \\
       115 & $  100 \pm    22$ & $   33 \pm     3$ & $  121 \pm     4$ & $   56 \pm     1$ & $   7 \pm 0.4$ &    10.0 & $  0.38 \pm   0.32$ & $  \phantom{-}0.09$ & $ -0.37$ \\
      116a & $  119 \pm    15$ & $   14 \pm     1$ & $   99 \pm     2$ & $   82 \pm     1$ & $   3 \pm 0.4$ &    20.3 & $  0.59 \pm   0.20$ & $ -0.08$ & $ -0.46$ \\
      116b & $  179 \pm    16$ & $   35 \pm     1$ & $  107 \pm     2$ & $   68 \pm     1$ & $   4 \pm 0.2$ &    22.3 & $  0.62 \pm   0.16$ & $ -0.22$ & $ -0.42$ \\
      116c & $  178 \pm    16$ & $  109 \pm     5$ & $  104 \pm     2$ & $   58 \pm     1$ & $  10 \pm 0.2$ &    17.0 & $  0.89 \pm   0.16$ & $ -0.23$ & $ -0.44$ \\
       117 & $  277 \pm    36$ & $  284 \pm    33$ & $   29 \pm     3$ & $   40 \pm     3$ & $  13 \pm 2.9$ &     1.1 & $  0.00 \pm   0.30$ & $ -0.97$ & $ -0.99$ \\
       118 & $  315 \pm    72$ & $  204 \pm    11$ & $   38 \pm     1$ & $   53 \pm     1$ & $   9 \pm 0.3$ &     3.9 & $  0.04 \pm   0.20$ & $ -0.91$ & $ -0.87$ \\
       121 & $  395 \pm   103$ & $  204 \pm    23$ & $   39 \pm     2$ & $   61 \pm     3$ & ... &     1.0 & $  0.13 \pm   0.44$ & $ -1.00$ & $ -0.86$ \\
       122 & $  268 \pm    54$ & $  167 \pm    15$ & $   31 \pm     1$ & $   55 \pm     2$ & $   7 \pm 0.5$ &     2.3 & $  0.43 \pm   0.34$ & $ -0.93$ & $ -0.96$ \\
       123 & $  319 \pm    47$ & $   39 \pm     5$ & $   34 \pm     3$ & $   82 \pm     4$ & ... &     1.4 & $  0.00 \pm   0.41$ & $ -0.97$ & $ -0.92$ \\
       124 & $  309 \pm    60$ & $  209 \pm    18$ & $   31 \pm     1$ & $   63 \pm     2$ & $   8 \pm 0.6$ &     1.5 & $  0.09 \pm   0.33$ & $ -1.00$ & $ -0.96$ \\
   128.225 & $  379 \pm    56$ & $   36 \pm     2$ & $   31 \pm     2$ & $  106 \pm     4$ & ... &     1.3 & $  0.00 \pm   0.25$ & $ -1.08$ & $ -0.96$ \\
       130 & $  274 \pm    20$ & $  146 \pm     6$ & $   19 \pm     1$ & $   44 \pm     2$ & $   6 \pm 1.7$ &     3.0 & $  0.00 \pm   0.18$ & $ -1.15$ & $ -1.17$ \\
   131.228 & $  378 \pm    82$ & $   62 \pm     6$ & $   29 \pm     1$ & $   76 \pm     3$ & ... &     4.4 & $  1.30 \pm   0.39$ & $ -1.10$ & $ -0.98$ \\
       203 & $  334 \pm    31$ & $  196 \pm    11$ & $   27 \pm     2$ & $   69 \pm     4$ & ... &     1.7 & $  0.00 \pm   0.29$ & $ -1.08$ & $ -1.01$ \\
       205 & $  181 \pm    25$ & $  397 \pm    17$ & $   17 \pm     2$ & $   23 \pm     3$ & $  14 \pm 1.2$ &     1.5 & $  0.00 \pm   0.21$ & $ -1.02$ & $ -1.21$ \\
       208 & $  357 \pm    38$ & $   64 \pm     4$ & $   47 \pm     3$ & $  155 \pm     7$ & ... &     1.3 & $  0.00 \pm   0.24$ & $ -0.88$ & $ -0.78$ \\
      210a & $  168 \pm    16$ & $   44 \pm     2$ & $   91 \pm     2$ & $   91 \pm     1$ & $   7 \pm 0.2$ &     8.2 & $  0.08 \pm   0.17$ & $ -0.26$ & $ -0.49$ \\
       213 & $  202 \pm     9$ & $   30 \pm     1$ & $  105 \pm     4$ & $   74 \pm     2$ & $   5 \pm 0.3$ &    18.9 & $  0.00 \pm   0.13$ & $ -0.28$ & $ -0.43$ \\
       214 & $  149 \pm     9$ & $   28 \pm     2$ & $   96 \pm     4$ & $   83 \pm     3$ & $   5 \pm 0.9$ &    14.4 & $  0.00 \pm   0.21$ & $ -0.19$ & $ -0.47$ \\
       215 & $  225 \pm    17$ & $   54 \pm     2$ & $  112 \pm     2$ & $  149 \pm     2$ & $   8 \pm 0.2$ &     5.7 & $  0.40 \pm   0.14$ & $ -0.30$ & $ -0.41$ \\
       216 & $  230 \pm    32$ & $  256 \pm    13$ & $   68 \pm     3$ & $  104 \pm     4$ & $  10 \pm 2.1$ &     2.6 & $  0.00 \pm   0.19$ & $ -0.53$ & $ -0.62$ \\
       217 & $  192 \pm     8$ & $   76 \pm     3$ & $   73 \pm     3$ & $   58 \pm     2$ & $   8 \pm 0.5$ &    31.7 & $  0.00 \pm   0.13$ & $ -0.42$ & $ -0.59$ \\
      221a & $  359 \pm    98$ & $  164 \pm    19$ & $   43 \pm     2$ & $   75 \pm     3$ & ... &     1.2 & $  0.27 \pm   0.44$ & $ -0.92$ & $ -0.82$ \\
      221b & $  353 \pm    43$ & $   75 \pm     6$ & $   31 \pm     4$ & $   84 \pm     6$ & ... &     0.9 & $  0.00 \pm   0.32$ & $ -1.05$ & $ -0.96$ \\
       223 & $  357 \pm    88$ & $   55 \pm     6$ & $   39 \pm     2$ & $   87 \pm     3$ & ... &     2.9 & $  0.58 \pm   0.39$ & $ -0.95$ & $ -0.86$ \\

\hline
 \end{tabular} }
 \end{center}
\raggedright{Line fluxes are normalized to H$\beta  = 100$.  The H$\beta$ flux in col.~7 is uncorrected for potential slit losses. }
\end{table*}

 We list the reddening-corrected line fluxes of our sample, in units of  H$\beta = 100$, in Table~\ref{tab:galres}. The quoted errors account for measurement, flux calibration and flat fielding errors.
We point out that the H$\beta$ flux (column~7) underestimates the total line flux (hence the ionizing luminosity, see \S\ref{sec:ionpop}), due to the unavoidable slit losses.

\begin{figure*}
   \centering
   \includegraphics[width = 8.8cm]{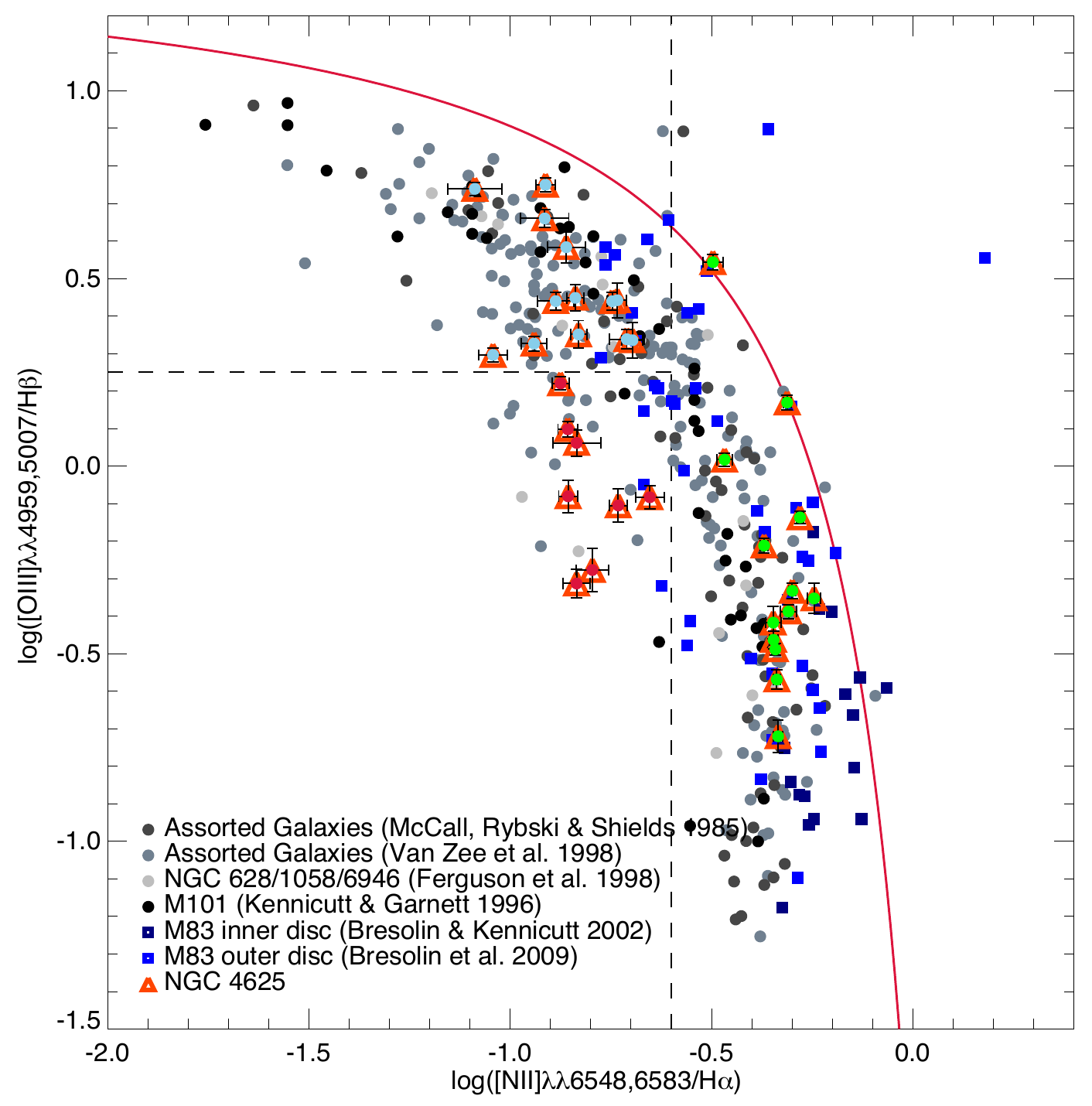} 
   \includegraphics[width = 8.8cm]{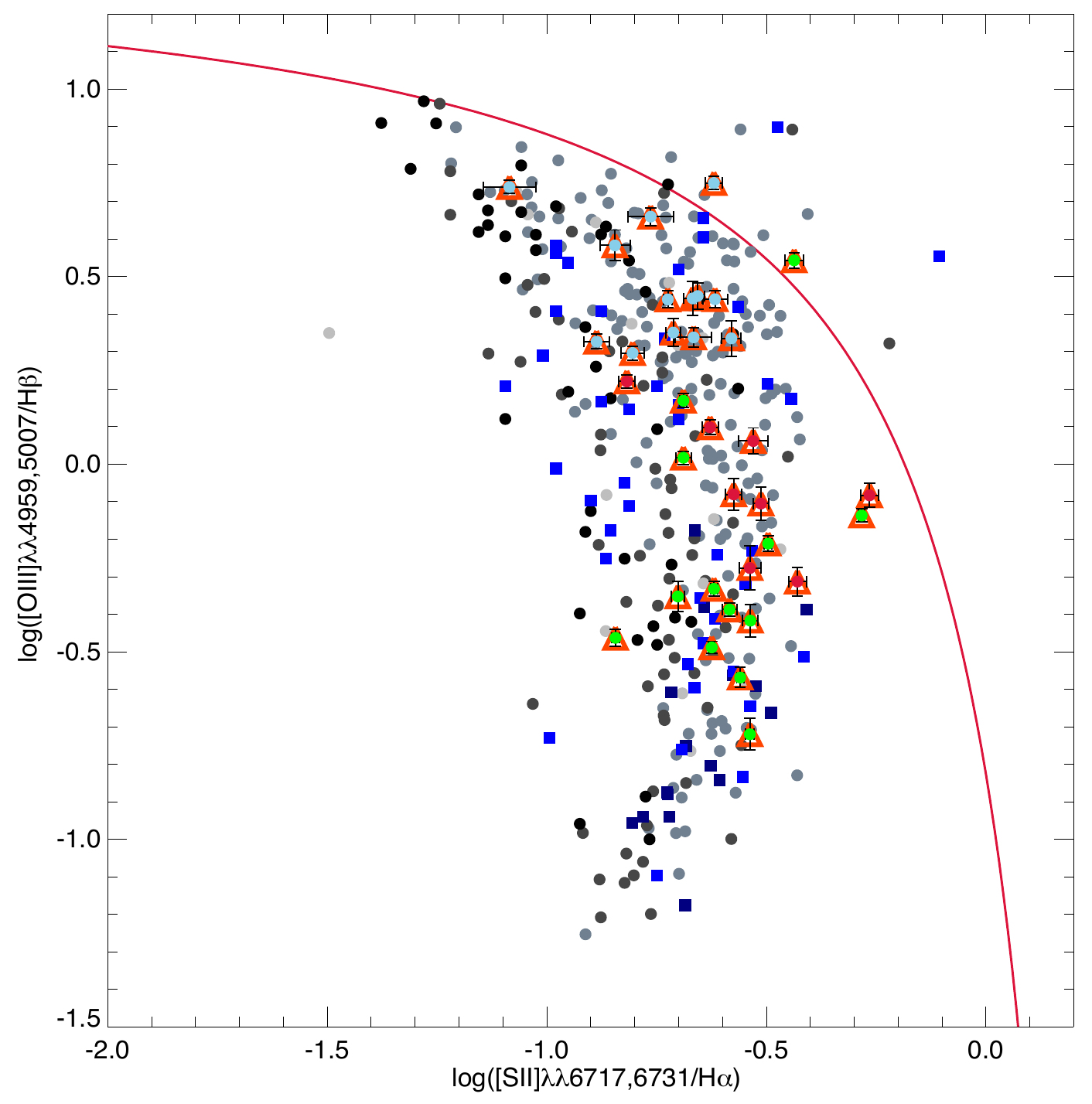} 
   \caption{BPT diagnostic plots. {\em Left:} \oiii/H$\beta$ vs.~\nii/H$\alpha$. {\em Right:} \oiii/H$\beta$ vs. \sii/H$\alpha$. Our sample of \hii\ regions in NGC~4625 (triangles) has been plotted together with published \hii\ region datasets. Objects in the inner disc  are further identified by green dots, while the outer disc \hii\ regions are marked either with cyan dots, indicating \oiii/H$\beta > 0.25$, or red dots for \oiii/H$\beta < 0.25$.
    The blue squares show the \hii\ regions observed in the other prototypical XUV galaxy, M83, with light blue squares representing the outer disc sample of \citet{bresolin09}, and dark squares the inner disc \hii\ regions of \citet{bresolin02}. The dots show previous studies of bright inner disc \hii\ regions, including the  samples of \citet[dark grey dots]{mccall85} and \citet[grey dots]{vanzee98}. The light grey dots represent the \hii\ regions in NGC~628, NGC~1058 and NGC~6946  studied by \citet{ferguson98a}. The M101 sample of \citet{kennicutt96} is represented by the black dots. 
   The red curves represent the upper boundaries for photoionised nebulae defined by \citet{kewley06}. The dashed lines in the left panel serve to schematically isolate objects that
   appear to deviate from the main ionisation sequence, towards the lower left of the diagram. 
   }
   \label{fig:bpt}
\end{figure*}

We have characterised our  \hii\ region sample using the familiar BPT \citep{baldwin81} diagnostic diagrams, by plotting in Fig.~\ref{fig:bpt} \nii/\ha\ against \oiii/H$\beta$ and \sii/\ha\ against \oiii/H$\beta$ \citep{veilleux87}. We represent our sample of \hii\ regions  with orange triangle symbols, and we include comparison data of mostly bright extragalactic \hii\ regions drawn from the literature: 

\begin{description}
	\item[{\em 1.} ] 99 \hii\ regions measured in 20 spiral and irregular galaxies by \citet[dark grey dots]{mccall85}. 
	\item[{\em 2.} ] 198 \hii\ regions in 13 spiral galaxies from \citet[grey dots]{vanzee98}. 
	\item[{\em 3.} ] 17 \hii\ regions from three late-type spirals (NGC~628, NGC~1058 and NGC~6946) studied by \citet[light grey dots]{ferguson98a}.
	\item[{\em 4.} ] 41 \hii\ regions in the spiral galaxy M101 observed by \citet[black dots]{kennicutt96}.
	\item[{\em 5.} ] 16 \hii\ regions from the inner disc of M83 from \citet[dark blue squares]{bresolin02}.
	\item[{\em 6.} ] 49 \hii\ regions from the M83 sample of \citet[light blue squares]{bresolin09}, mostly composed of outer disc \hii\ regions. Note that the two objects which appear above the theoretical boundary  in Fig.~\ref{fig:bpt} were identified as supernova remnants. 
\end{description}

\hii\ regions in the inner disc of NGC~4625 are further identified by green dots, while the outer disc \hii\ regions are marked either with cyan dots, indicating \oiii\llin4959,\,5007/H$\beta > 0.25$, or red dots (\oiii\llin4959,\,5007/H$\beta < 0.25$). This arbitrary distinction helps us to distinguish these two subsets. In fact, in the \nii/\ha\ vs. \oiii/H$\beta$ diagram not all of the inner disc \hii\ regions  sit on the tight sequence defined by bright \hii\ regions:
those marked with red dots appear to lie below and/or to the left of the main ionisation sequence. On the other hand, the \sii/\ha\ versus \oiii/H$\beta$ plot shows a similar ionisation sequence for objects in the inner and extended disc of NGC~4625, which is consistent with the ionisation sequence defined by bright extragalactic \hii\ regions.
These results will be further discussed in \S \ref{sec:roguehii}.

 Eight of our \hii\ regions  were also studied by G07 (identified in Table~\ref{tab:posinfo} by the {\sc xuv} numbers used in their paper). The H$\beta$ fluxes measured by G07 are systematically higher than ours by a mean factor of 1.45, as 
expected from their 50\% larger slit widths (1\farcs5 \vs\ 1\farcs0), and thus it appears that the poor seeing conditions that we experienced
did not significantly affect our flux measurements.
Our \oiii/H$\beta$ ratios for 6 \hii\ regions in common (not all of the G07 had a measured \oiii\ line) are systematically 22\% lower, on average, for which we do not have an explanation.
Of the only four targets for which the intensity of the \oii\lin3727 line could be compared, we found good agreement in two cases (our regions 118 and 122), a 34\% lower flux ratio in a third (region 103), which is mostly 
due to our lower assumed reddening correction, while in the case of region 130 our  \oii/\hbeta\ value is almost a factor of 2 lower. In this last case, the reddening correction cannot explain the large discrepancy.
It is well-known (e.g.~\citealt{bresolin05}) that the \oii\lin3727 line is  susceptible to systematic errors in nebular optical  spectra, as a result of the increased uncertainty in the flux calibration of the blue end of the spectrum, smaller throughput at these wavelengths, and errors in the reddening corrections. We conclude by noting that most of the spectra shown in Fig.~4 of G07 are very noisy in the region around 4000\,\AA, and in several cases even the \hbeta\ line seems to be detected with a poor signal-to-noise ratio (unfortunately G07 did not publish error estimates for their line ratios, so this statement is based on a visual assessment of their data). Thanks to longer exposure times (2 hours \vs\ 40 minutes) with a larger telescope aperture (Subaru 8\,m \vs\ Palomar 5\,m), our spectra are, as should be expected, of superior quality.
We are therefore confident in the reliability of our emission line data and the estimated uncertainties presented in Table~\ref{tab:galres}. 


\section{Nebular Abundances}\label{sec:nebabunds}

\begin{table*}
 \caption{Auroral line-derived abundances.}
 \label{tab:te}
 \begin{tabular}{@{}l c c c c}
  \hline
  \hline
  ID & \oiii\lin4363/H$\beta$ & \te\ (K) & 12\,+\,log(O/H) & log(N/O) \\
  (1) & (2) & (3) & (4) & (5) \\ 
  \hline
  104.204 & $2.0\pm1.5$ & $14000 \pm 2700$ & $7.95\pm0.32$ & $-1.09\pm0.34$ \\
  108 & $8.2\pm1.4$& $15300 \pm 1100$ & $8.01\pm0.09$ & $-1.27\pm0.12$\\
  205 & $7.7\pm4.2$& $15100 \pm 2500$ & $7.93\pm0.25$ & $-1.35\pm0.31$\\
  \hline
\end{tabular} 
\end{table*}

\subsection{Direct Abundances}

Ideally, the derivation of the chemical abundances of ionised nebulae requires first a measurement of  the electron temperature \te, since the line emissivities strongly depend on it. The classical method to directly obtain \te\ involves the simultaneous measurement of the auroral \oiii\lin4363 line and of the nebular \oiii\llin4959,\,5007 lines. This proves to represent a challenge with the increase of the metallicity, because  the cooling mechanisms become highly efficient and \oiii\lin4363 too weak to observe. However, having a few \lin4363 detections is important, since they allow the derivation of nebular abundances that are independent of the systematic uncertainties of the strong line methods described below.
In our NGC~4625 sample  three \hii\ regions were found to have a reliable \oiii\lin4363 detection. With routines in the {\sc iraf \em nebular} package we  obtained \te\ and subsequently the O/H and  N/O ratios, assuming that N/O = N$^{+}$/O$^{+}$. These quantities are summarised in Table~\ref{tab:te}.

\subsection{Strong Line Abundances} \label{sec:slabunds}

Without detection of the temperature-sensitive \oiii\lin4363 line we must resort to strong line methods to determine the oxygen abundances. Several such methods, using a variety of line ratios, are present in the literature, with calibrations that are provided either empirically or by theoretical considerations. Unfortunately, it is well established that the various methods provide differing solutions for the chemical composition of star forming galaxies and \hii\ regions, resulting in
 systematic shifts  of up to $\sim$0.7 dex  \citep{kewley08, bresolin09b}.
By providing the reader with results from different  strong line indicators we hope to detect any robust trends in the abundance gradient of NGC~4625, and at the same time highlight the limitations of our findings. Although these indicators were originally developed for the study of the chemical abundances of luminous extragalactic  \hii\ regions, tests on Galactic nebulae ionised by single or only a few stars have excluded the presence of systematic effects \citep{kennicutt00,oey00}. Further empirical tests were carried out by our team when analysing a set of faint outer disc \hii\ regions in M83 \citep{bresolin09}.\\

We have considered the following strong line abundance indicators in the remainder of our work:

\begin{description}
	\item[{\rm 1.} ] R$_{23}$ = (\oii\lin3727 + \oiii\llin4959,\,5007)/H$\beta$. This is one of the most widely used strong line abundance indicators, however it does have some limitations \citep[e.g.][]{perez-montero05}. In particular, the R$_{23}$ indicator 
has a double valued nature, peaking with a value log(R$_{23}$)\,$\simeq$\,1 around
\oh\ = 8.5 in the theoretical calibration provided by \citet{mcgaugh91}. Smaller R$_{23}$ values correspond to
both lower and higher O/H abundances. The turnover occurs at lower abundances, \oh\,$\simeq$\,8.0,
when using O/H empirical determinations from \oiii\lin4363 detections (\citealt{bresolin07,bresolin08}).
Because of this degeneracy, it is essential to establish whether the \hii\ regions under scrutiny fall into the lower or upper R$_{23}$ branch. This is commonly accomplished with the help of line ratios that vary monotonically with metallicity, such as \nii/\ha\ and \nii/\oii. \\[-2mm]
	
	\item[{\rm 2.} ] N2 = \nii\lin6583/H$\alpha$, as empirically calibrated by \citet{pettini04}. 	N2 is a monotonic function of the oxygen abundance and thus avoids the degeneracy problem associated with R$_{23}$. However, \citet[\,=\,KD02]{kewley02} showed that  N2 is sensitive to the ionisation parameter.\\[-2mm]
	
	\item[{\rm 3.}] N2O2 = \nii\lin6583/\oii\lin3727, a useful indicator because it is highly insensitive to the ionisation parameter \citep{dopita00}. We have used two separate calibrations in this study, the theoretical one by KD02, based on grids of photoionisation models, and the empirical one by \citet[\,=\,B07]{bresolin07}, based on a sample of \hii\ regions with available \te\ abundances.
\end{description}

We note that N2 and, to a lesser extent, R$_{23}$ are more robust against uncertainties in the atmospheric extinction correction compared to the N2O2 indicator,  which is formed  by emission lines spread across a very wide spectral range. N2, in particular, is also virtually insensitive to
uncertainties in the interstellar reddening correction and to the potential effects of variable seeing, which could be significant
for flux ratios that involve emission lines measured from separately obtained blue and red spectra, as is often the case for \nii/\oii.

\subsubsection{Results}

\begin{figure*}
   \centering
   \includegraphics[width = 16cm]{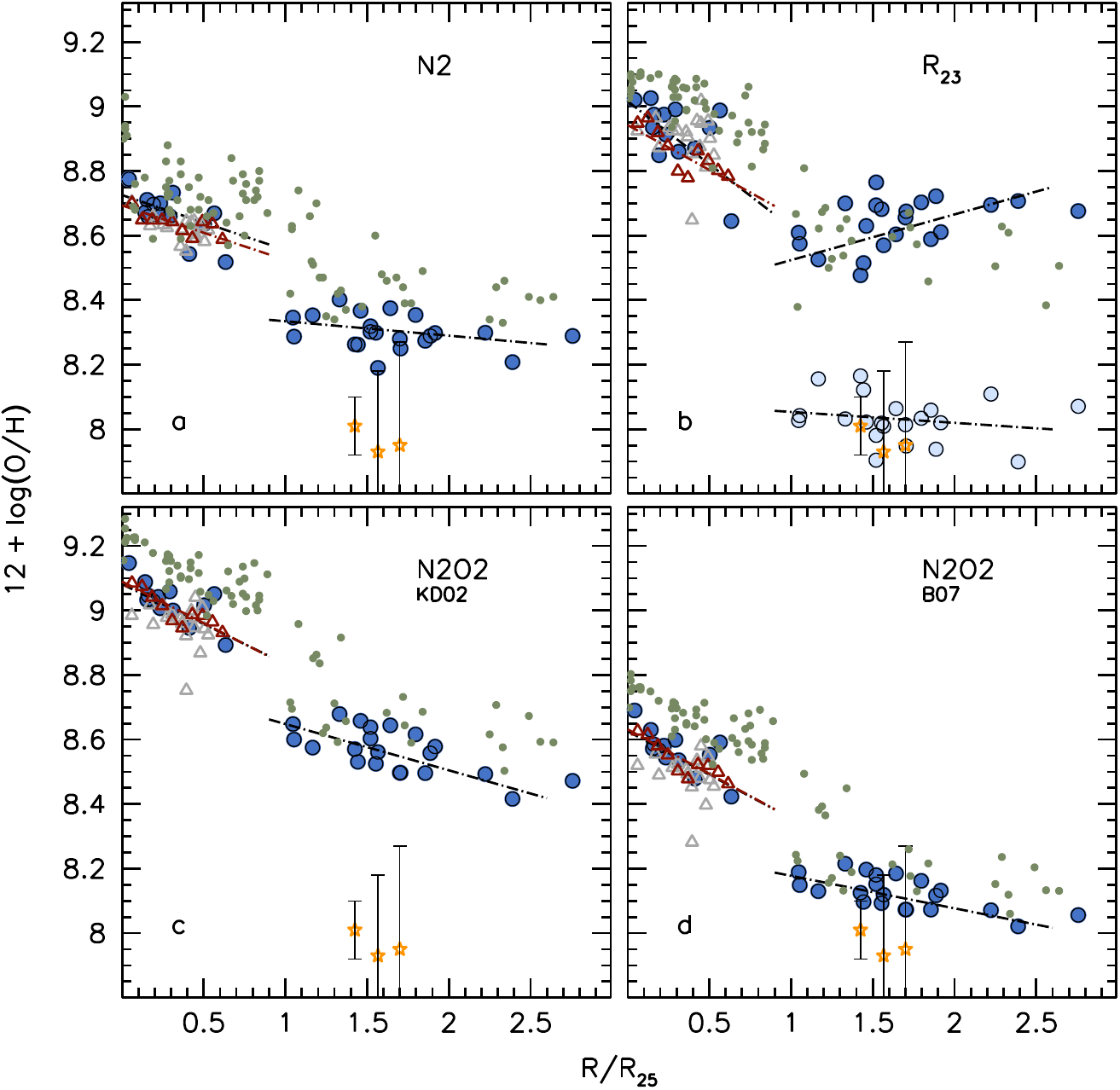} 
   \caption{The radial oxygen abundance gradient in NGC~4625 (coloured circles), determined using various strong line methods: (a) N2, calibrated by \citet{pettini04}; (b) R$_{23}$, with upper branch calibration shown in blue, and lower branch calibration in light blue; (c) N2O2, theoretical calibration by \citet{kewley02}; (d) N2O2, empirical calibration by \citet{bresolin07}. \te-based abundances have been added to all four panels as gold star symbols. We include the results obtained  from measurements of the inner disc from the PINGS survey, including individual \hii\ regions studied in our sample (grey triangles) and radially averaged abundances (red triangles). For comparison, the M83 data points from \citet{bresolin09} are shown by the green dots. The lines represent weighted least square fits to our Subaru data for the inner and outer disc (black) and to the 
radially averaged PINGS data (red).  }
   \label{fig:abund}
\end{figure*}

\begin{table*}
 \caption{Coefficients of least-squares fitting to the relation \oh\,=\, $a - b$\,(R/R$_{25}$).}
 \label{tab:fits}
 \begin{tabular}{@{}l c c c c c}
  \hline
  \hline
 & Abundance & \multicolumn{2}{c}{Inner Disc} & \multicolumn{2}{c}{Outer Disc} \\
& Indicator & $a$ & $b$ & $a$ & $b$  \\
  \hline
  Our Sample	& N2 & 	$8.72 \pm 0.03$ & $-0.17 \pm 0.08$ & $8.38 \pm 0.03$ & $-0.05 \pm 0.02$ \\
    	& R$_{23}$ (upper branch)& 	$9.02 \pm 0.04$ & $-0.39 \pm 0.13$ & $8.38 \pm 0.08$ & $+0.14 \pm 0.05$ \\
     	& R$_{23}$ (lower branch)& 	-- & -- & $8.09 \pm 0.07$ & $-0.03 \pm 0.03$ \\
     	& N2O2 (KD02) & 	$9.08 \pm 0.02$ & $-0.25 \pm 0.08$ & $8.79 \pm 0.06$ & $-0.14 \pm 0.04$ \\
        & N2O2 (B07) & 	$8.63 \pm 0.02$ & $-0.27 \pm 0.08$ & $8.28 \pm 0.04$ & $-0.10 \pm 0.03$ \\
  \hline
PINGS  & N2 & $8.69 \pm 0.01$ & $-0.17 \pm 0.03$ & -- & -- \\
(Radial data)  	& R$_{23}$  (upper branch) & 	$8.94 \pm 0.04$ & $-0.28 \pm 0.08$	& -- & -- \\
    	& N2O2 (KD02) & 	$9.09 \pm 0.02$ & $-0.26 \pm 0.04$ & -- & -- \\
      	& N2O2 (B07) &  	$8.63 \pm 0.02$ & $-0.27 \pm 0.05$ & -- & -- \\
	\hline
\end{tabular}
\end{table*}

In Fig.~\ref{fig:abund} we present the abundances we derived from the strong line indicators, plotted as a function of galactocentric distance.
The three targets with \te-based abundances appear as gold stars. We show with black dot-dashed lines the weighted least squares fits to the data for both the inner (R\,$<$\,R$_{25}$) and outer (R\,$>$\,R$_{25}$) disc.
We also include as a comparison the sample of M83 \hii\ regions studied by \citet[green dots]{bresolin09}, further discussed in \S\ref{sec:reveal}.

The triangles in Fig.~\ref{fig:abund}  represent measurements of \hii\ regions in NGC~4625 from PINGS (PPAK Integral Field Spectroscopy Nearby Galaxy Survey,  see \citealt{rosales10}), which obtains spectra from the PMAS integral field unit at the Calar Alto 3.5\,m telescope. To compare the PINGS results to our own we followed two different approaches, either by  
averaging spectra from all fibers contained within equally spaced ($4''$ separation) radial  bins (red triangles in Fig.~\ref{fig:abund}), or by using the 
sum of individual fibers  covering the spatial extent of individual \hii\ regions (grey triangles).
The agreement between the abundances derived from PINGS and our data is excellent, as shown by the close match between the  regressions to the average radial bins (red lines) and to our inner disc results (black). This can be quantitatively appreciated by looking at the coefficients of the weighted linear fits to the the radial trends of metallicity that are summarized in Table~\ref{tab:fits}.

\medskip
Following is a brief discussion of the results obtained from the individual abundance diagnostics:\\[2mm]
\noindent
{\em R$_{23}$ --}
The interpretation of the R$_{23}$-based  abundances (top right-hand panel of Fig.~\ref{fig:abund}) requires some important considerations.
All the inner disc \hii\ regions in NGC~4625 have log\,(\nii/\ha)\ $>$ $-0.8$, which, according to the KD02 photoionisation models, corresponds to  the upper branch of R$_{23}$. The result could be  more ambiguous for the  outer disc \hii\ regions, which are  closer to the turnover region in the R$_{23}$ \vs\ metallicity relationship, even though virtually all lie in the upper-branch regime, according to the \citet{kewley08} criterion, log\,(\nii/\ha)\ $>$ $-1.1$. We reach the same conclusion from the  log\,(\nii/\oii) values (ranging between $-1.1$ and $-0.7$).
Therefore, we used  the upper branch solution for the inner disc, 
whilst for the outer disc  we used both the lower branch (light blue) and the upper branch (dark blue) calibrations of \citet[in the analytical form given by \citealt{kobulnicky99}]{mcgaugh91} for comparison.

Taking at face value the information obtained from the \nii/\ha\ and \nii/\oii\ line ratios we would be led to favour the upper branch solution for the whole sample of \hii\ regions, leading, in particular, to  \oh\,$\sim$\,8.6 in the outer disc. However, this is in evident contrast with the
 direct abundances that we measured for the three outer disc targets presented in Table~\ref{tab:te}.
 Their oxygen abundance, \oh\ $\sim$ 8.0, agrees very well with the {\em lower} branch solution shown in Fig.~\ref{fig:abund}. This inconsistency is the result of the still unresolved discrepancy between theoretical and direct calibrations of strong-line indicators. We note that our measurements for the direct abundances  essentially fall in the transition range \oh\,=\.8.0 - 8.25
 between upper and lower branch of the R$_{23}$ calibration based on the P-parameter of \citet{pilyugin2005}, which is directly tied to the empirical, \oiii\lin4363-based oxygen abundances. Consequently, we cannot
 obtain a reliable estimate of the outer disc abundances from this method, although clearly they would be systematically lower than those from the theoretical \citet{mcgaugh91}.

\medskip
\noindent
{\em N2 --} We point out that this diagnostic, when applied to the three targets with a \lin4363 detection, provides abundances that are 0.3 dex higher relative to the direct method (top left-hand panel of Fig.~\ref{fig:abund}). This is not fully expected, since the N2 calibration by \citet{pettini04} relies mostly on extragalactic \hii\ regions where the oxygen abundances had been derived from auroral line detections However, a similar effect was observed in M83 by \citet{bresolin09}.
We can speculate on the possible causes of this systematic difference, pointing out that the the ionization properties of the calibration sample
can differ significantly from those of the faint  nebulae found in the extended discs, and that the N2 diagnostic is sensitive to variations in the ionization parameter (KD02).

\medskip
\noindent
{\em N2O2 --}
The lower right-hand panel of Fig.~\ref{fig:abund} shows the strong line abundances calculated with the calibration of the \nii/\oii\ \vs\ O/H relation by \citet{bresolin07}. Unsurprisingly, this  indicator provides a good match, within the uncertainties, to the direct \te-based abundances,  since the calibration  was empirically derived from extragalactic \hii\ regions with auroral line detections. 
The well-known systematic discrepancies associated with the use of different strong line indicators and their calibration are evident in Fig.~\ref{fig:abund}. The theoretical N2O2 calibration of \citet{kewley02} yields abundances that are roughly 0.45 dex higher than the empirical calibration of \citet{bresolin07}. Similar systematic shifts are present between N2 and N2O2.

\subsubsection{What do the Strong Line Abundances Reveal?}\label{sec:reveal}

Despite the systematic offsets between different abundance diagnostics noted above, 
the results summarised in Fig.~\ref{fig:abund} appear to be qualitatively consistent with each other, in particular showing a shallower gradient in the outer disc, as well as an abundance discontinuity between the inner and outer disc, occurring near R$_{25}$. Extrapolation of the inner and outer disc abundance gradients suggests a discontinuity of approximately 0.15-0.2 dex, based on the N2 and N2O2  abundance indicators. As Table~\ref{tab:fits} shows, the slopes of the gradients depend somewhat on the choice of abundance indicator.
In the inner disc, N2 produces the flattest gradient ($-0.17$\,dex/R$_{25}$), about half of the value obtained from R$_{23}$ ($-0.39$\,dex/R$_{25}$). The scatter of the data points is, however, quite large in the R$_{23}$ case. If we look at the radially averaged PINGS data we find a good agreement between R$_{23}$ and N2O2 ($\sim$ $-0.27$\,dex/R$_{25}$), still slightly larger (2\,$\sigma$) than the slope provided by N2. The outer disc slopes are also smaller (1\,$\sigma$) for N2 ($-0.05$\,dex/R$_{25}$) than for N2O2 ($-0.10$ to $-0.14$ \,dex/R$_{25}$). This agrees with the conclusion by \citet{bresolin09b} in their detailed study of NGC~300 that abundance gradient slope values have a dependency on the method adopted to measure them, contrary to 
other statements in the literature, that changing the strong-line abundance calibration only affects the zero points.

We  include in Fig.~\ref{fig:abund} the data on the inner and outer \hii\ regions in M83 analyzed by \citet[green dots]{bresolin09}. This allows us to compare the results we obtained for NGC~4625 to the only other extended disc galaxy studied in detail until now. As the figure shows, there is a very good qualitative match between the two galaxies. The change in the galactocentric abundance slope between inner and outer disc is comparable, as is the discontinuity in O/H detected around the isophotal radius, as well as the slope of the gradients themselves. The 
outer discs in the two galaxies have remarkably similar oxygen abundances, regardless of the metallicity indicator used for the comparison. The inner disc abundances in M83 are $\sim$0.15 dex higher than in the case of NGC~4625, which is, at least qualitatively, expected from the higher absolute luminosity of the former ($M_B=-20.8$ \vs\ $M_B=-17.2$ adopting distances of 4.5 and 9.5~Mpc, respectively). Panel (b) of Fig.~\ref{fig:abund} also suggests that the apparent positive abundance gradient obtained in the outer disc of NGC~4625 from the use of the upper branch of R$_{23}$ could be due to small number statistics at large radius, since the scatter of the points is similar for the two galaxies, and M83 does not show a positive gradient in its outer parts. This would imply that the R$_{23}$-inferred change of slope between inner and outer disc of NGC~4625 would be made roughly consistent with what is observed with the other diagnostics.

\section{Analysis of the H\,{\small II} Region Sample} \label{sec:roguehii}

Earlier on (Fig.~\ref{fig:bpt}) we showed the familiar BPT diagnostic diagrams for our sample in NGC~4625, concluding that all objects  were indeed photoionized nebulae. However, these plots also revealed a subset of outer disc \hii\ regions which apparently lie off the main ionisation sequence defined by extragalactic giant \hii\ regions in the \nii\ vs. \oiii\ plane, whilst apparently conforming to the less robust sequence formed in the \sii\ vs. \oiii\ plane.

\subsection{Properties of H\,{\small II} Regions in the Outer Disc } \label{sec:hiianalysis}

We have arbitrarily subdivided the outer disc \hii\ regions in NGC~4625 by making a cut at \oiii/H$\beta = 0.25$, represented in Fig.~\ref{fig:bpt} with a horizontal dashed line. The outer \hii\ regions which lie on the tight, well defined ionisation sequence shown in the left panel of Fig.~\ref{fig:bpt} fall above our cut and are highlighted with central cyan dots. Those  which lie off the ionisation sequence and below our \oiii/\hbeta\ cut were marked with red dots. It is not obvious whether the latter deviate as a result of either a smaller \nii\ or a smaller \oiii\ emission (or both) relative to the sequence defined by bright \hii\ regions.  We note that a few of the outer disc objects in M83, as well as a small number of nebulae in other galaxies, also lie below and to the left of the ionisation sequence.

\begin{figure}
   \centering
   \includegraphics[width = 8.8cm]{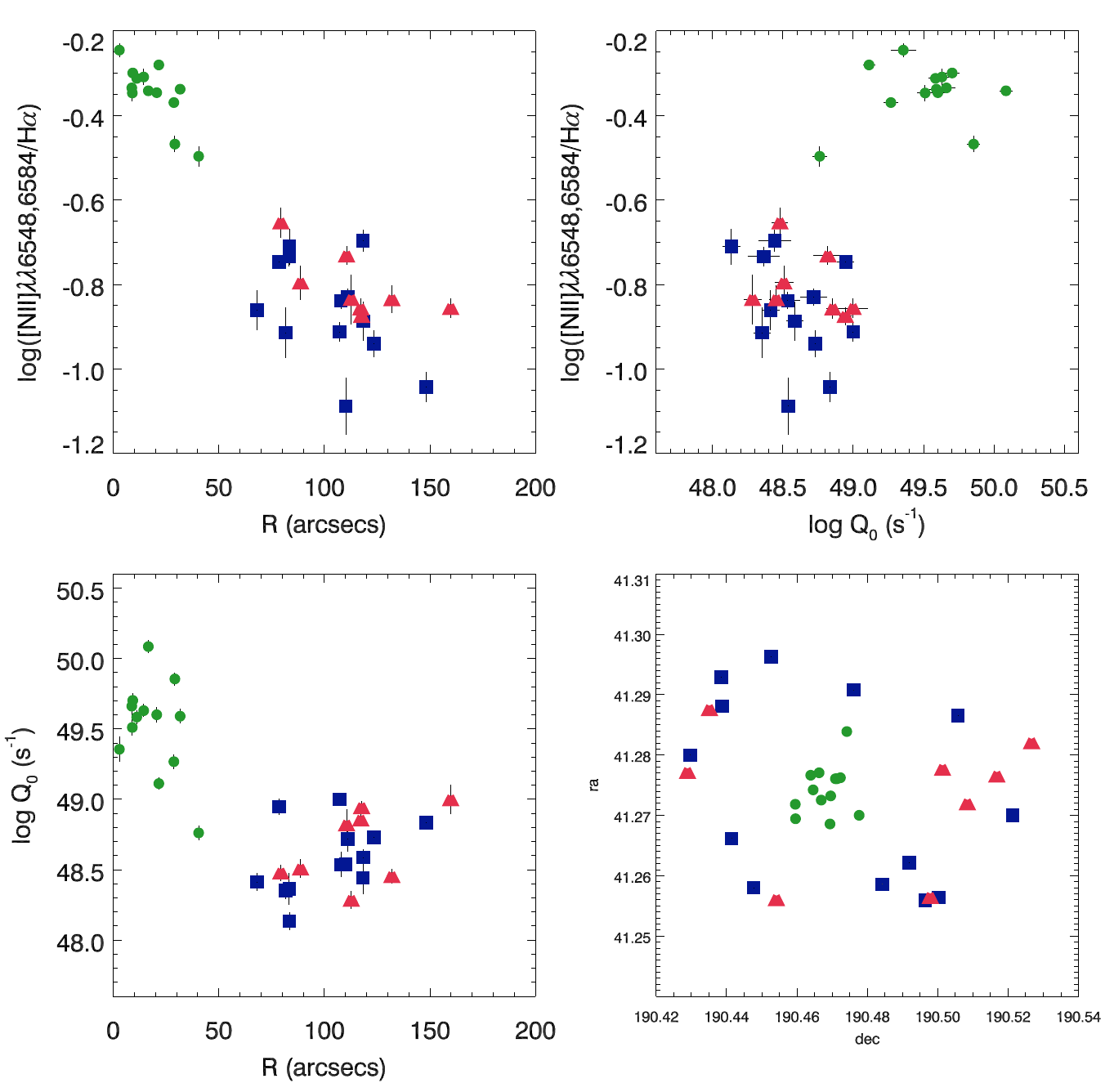} 
   \caption{Properties of outer disc \hii\ regions, for objects with \oiii/H$\beta > 0.25$ (blue squares) and \oiii/H$\beta < 0.25$ (red triangles) marked separately.
   Top-left: \nii/\ha\ vs. galactocentric distance. Top-right: \nii/\ha\ as a function of  ionising luminosity. Bottom-left: ionising luminosity as a function of galactocentric distance. Bottom-right: spatial distribution of individual \hii\ regions. Inner disc \hii\ regions are indicated by the green dots. }
   \label{fig:hiicross}
\end{figure}

It is worth testing whether these apparently deviant \hii\ regions share any common attributes, such as ionising luminosity, spatial distribution, metallicity and so forth. This is done in Fig.~\ref{fig:hiicross}. The top-left panel shows the \nii/\ha\ line ratio, i.e.~the N2 chemical abundance indicator, as a function of galactocentric radius. No differential behaviour between the two groups of outer disc \hii\ regions (blue and red symbols) is evident. The two groups  also  appear to share similar ionising luminosities (upper right and lower left panels) and to be  evenly distributed throughout the XUV disc of NGC~4625, without the presence of obvious separate sub-structures (lower right). There are slight indications that the deviant \hii\ regions have a systematically higher \nii/\ha\ line ratio than the other outer disc \hii\ regions, but this could easily be explained by the variance of the two distributions. 

\subsection{Systematics in the Diagnostic Diagrams}

In the following we consider what might affect the line fluxes  in such a way as to produce the observed deviation of a small number of outer disc \hii\ regions from the main ionisation sequence in the BPT diagnostic diagram.

\subsubsection{Stochastic Variations} \label{sec:ionpop}

\begin{figure}
   \centering
   \includegraphics[width = 8.3cm]{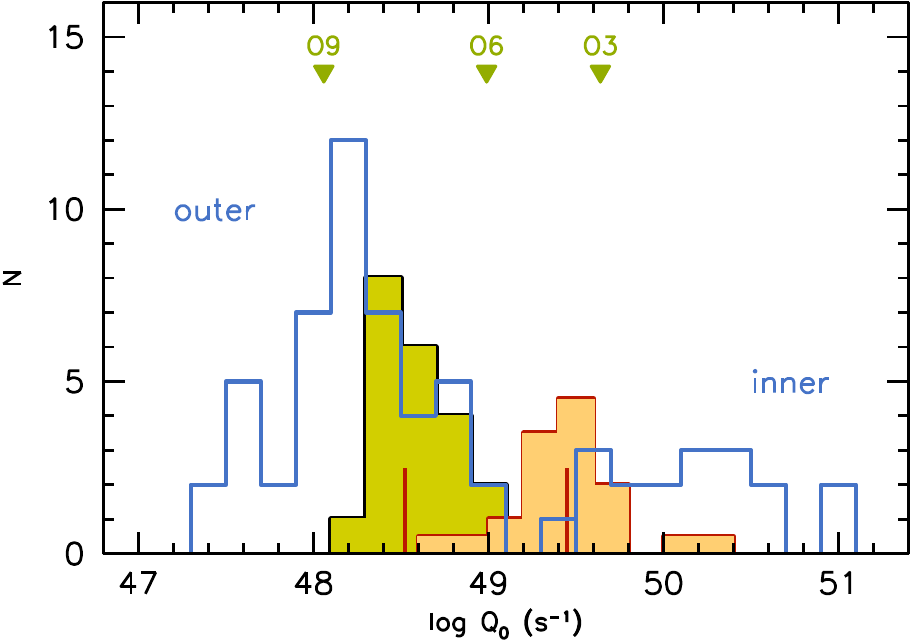} 
   \caption{Histograms of the H ionising photon flux, $\rm Q_0$, of \hii\ regions in our sample, uncorrected for slit losses. The green and orange shaded areas represent \hii\ regions in the outer and inner disc of NGC~4625, respectively. We have marked the median values for the two populations (log\,$\rm Q_0$\,=\,48.52 and 49.45, respectively). The blue lines show the histogram of the M83 sample studied by \citet{bresolin09}.
   The triangles at the top represent reference $\rm Q_0$ values for single hot stars (O3, O6 and O9), from \citet{martins05}.}
   \label{fig:hiiionhist}
\end{figure}

The direct observation of the ionising population of nebulae in outer galaxy discs are extremely difficult, thus the properties of the  associated stellar clusters have so far been obtained from studies of their  \ha\ and UV luminosities \citep{gildepaz05, gildepaz08, goddard10}, together with their IR emission \citep{dong08}. The total ionising photon flux of individual clusters can be easily estimated from the \ha\ luminosity,  using the conversion given by \citet{kennicutt98}. In Fig.~\ref{fig:hiiionhist} we show histograms of the hydrogen ionising photon flux for both the inner (orange) and outer (green) disc \hii\ regions, along with their median values. As a reference, we also show the hydrogen ionising fluxes of O3, O6 and O9 dwarf stars, taken from \citet{martins05}. The luminosities that we measure do not account for slit losses, which are expected to be smaller for the outer disc \hii\ regions, given that a greater percentage of their flux will fall through the 1$''$ slits compared to the brighter and larger inner disc \hii\ regions. We  estimated the slit losses by measuring \hii\ region fluxes  on an \ha\ image of NGC~4625 provided by the SINGS survey \citep{kennicutt03}. In the case of  the faintest \hii\ regions  losses are in the  5-10\% range, while for the  brighter inner \hii\ regions they increase to 20-70\%.

Figure \ref{fig:hiiionhist} confirms what other authors already found, namely that that the inner disc sample of \hii\ regions is brighter than the outer disc population by at least one order of magnitude (more if slit losses were considered).  
The ionising flux of typical outer disc \hii\ regions corresponds to that of a single late-type O star
(see \citealt{gildepaz05, thilker05, dong08,goddard10}). Consequently, it is necessary to  account for the stochastic nature
of the upper Initial Mass Function (IMF) of the ionizing clusters.
An ionising population drawn from a stochastically sampled IMF, as opposed to a fully sampled one, will tend to have a lower
 effective temperature, due to the preferential depletion of the rarest, most massive main sequence stars. This has important consequences for
 the total ionizing flux output, as well as for the hardness of the ionizing radiation.
In particular, as a result of softer spectral energy distributions, line ratios involving both high- and low-ionization potential species
could be significantly affected.

\subsubsection{Age effects} \label{sec:hiiage}

As the ionizing population of an \hii\ region ages the most massive stars die first, effectively truncating the mass function of the remaining stellar population. Not only could this alter the emission line fluxes, but we can also think of this effect as a proxy for stochastic variations which, as we have just seen, also act in the send of removing the high mass end of the stellar population. 

\begin{figure}
   \centering
   \includegraphics[width = 8.8cm]{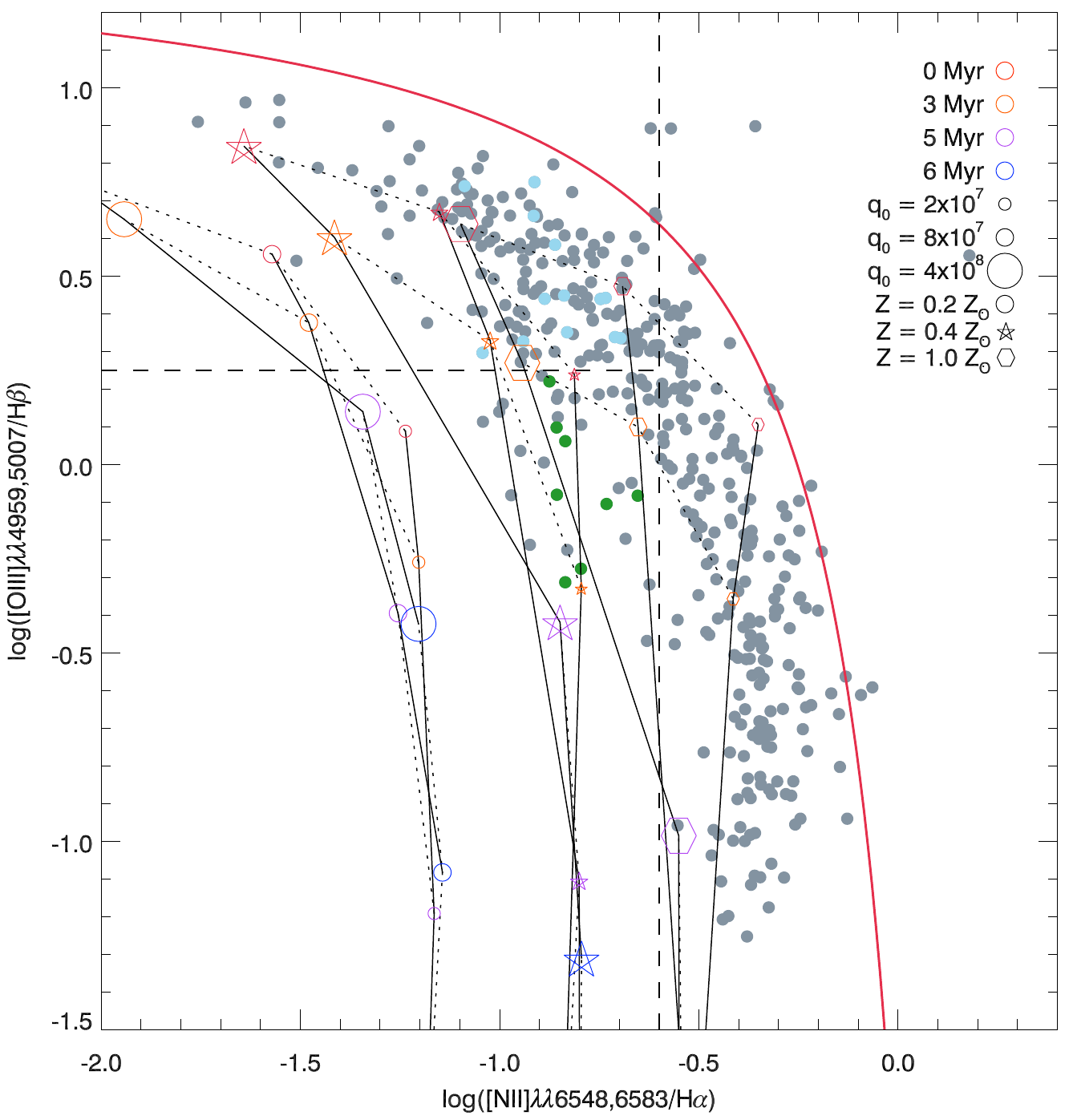} 
     \caption{Models in the \oiii/\hbeta\ \vs\ \nii/\halpha\  diagnostic diagram. The grey dots represent the same extragalactic \hii\ region samples included in previous figures.  Cyan symbols are used for  outer disc \hii\ regions in NGC~4625 for which \oiii/H$\beta > 0.25$, whilst green dots are used for those with \oiii/H$\beta < 0.25$. The red line represents the upper boundary for photoionised nebulae defined by \citet{kewley06}.  Black lines link models with identical  ionisation parameter, whilst the dotted lines link models of the same  age.}
   \label{fig:mod1}
\end{figure}

We make a simple attempt to model  such age effects by using the {\sc itera} program \citep{groves10}, along with the starburst models of \citet{levesque10}, which adopt a Salpeter IMF and the Starburst99 (\citealt{leitherer99}) stellar spectral energy distributions with standard mass loss rates. In Fig.~\ref{fig:mod1} we show a set of models on the \oiii/\hbeta\ \vs\ \nii/\halpha\ diagnostic diagram, calculated for a range of metallicities (Z = 0.2, 0.4, and 1.0 \zsun),  ionisation parameters (Q$_{0}$ = $2\times10^7$, $8\times10^7$, and $4\times10^8$) and ages (t~=~0, 3, 5, 6 Myr). The \hii\ region samples introduced in \S \ref{sec:obs} are represented by the grey dots, whilst our targets in  NGC~4625 are shown either with cyan (\oiii/H$\beta > 0.25$) or green (\oiii/H$\beta < 0.25$) symbols.

The zero-age models in Fig.~\ref{fig:mod1} lie on top of the ionisation sequence defined by the main extragalactic sample of giant \hii\ regions. The models show that, as the \hii\ regions evolve with time after the initial burst of star formation,
the \oiii\ emission rapidly decreases, and the good match with the observational data quickly disappears (see earlier results by \citealt{bresolin99}).
 The models also depart from the main ionisation sequence to lower \nii/\ha\ values as the metallicity
 decreases. The best fit to our deviant \hii\ regions (green dots) is obtained with the 0.4~\zsun\ models and ages between 3 and 5 Myr, depending on the ionisation parameter. 

The model comparison offers some clues as to why  some \hii\ regions  appear to deviate from the main ionisation sequence. Our interpretation relies on the fact that, as mentioned earlier, the evolution of the ionising output of low-luminosity \hii\ regions with age can mimic the effects due to stochastic variations in the number of
ionizing stars.
One might ask why so few \hii\ regions have been seen with similar properties in previous studies. The reasons are two-fold. Firstly, most of the previous studies have focused on inner disc \hii\ regions, which tend to have higher metallicities. According to our models, these objects would be nearly superimposed onto the main ionisation sequence. Secondly, there is a selection effect at work, since in the inner discs the brightest and youngest \hii\ regions are preferentially studied, also as a result of source confusion and crowding near the central regions  and along the spiral arms, while in the outer disc we attempt to sample as many \hii\ regions as possible, even the faintest ones, thus probing to older ages. This conclusion could be tested with  spectroscopic observations of larger samples of inner  disc \hii\ regions, reaching to fainter emission levels.

 \begin{figure}
   \centering
   \includegraphics[width = 8.8cm]{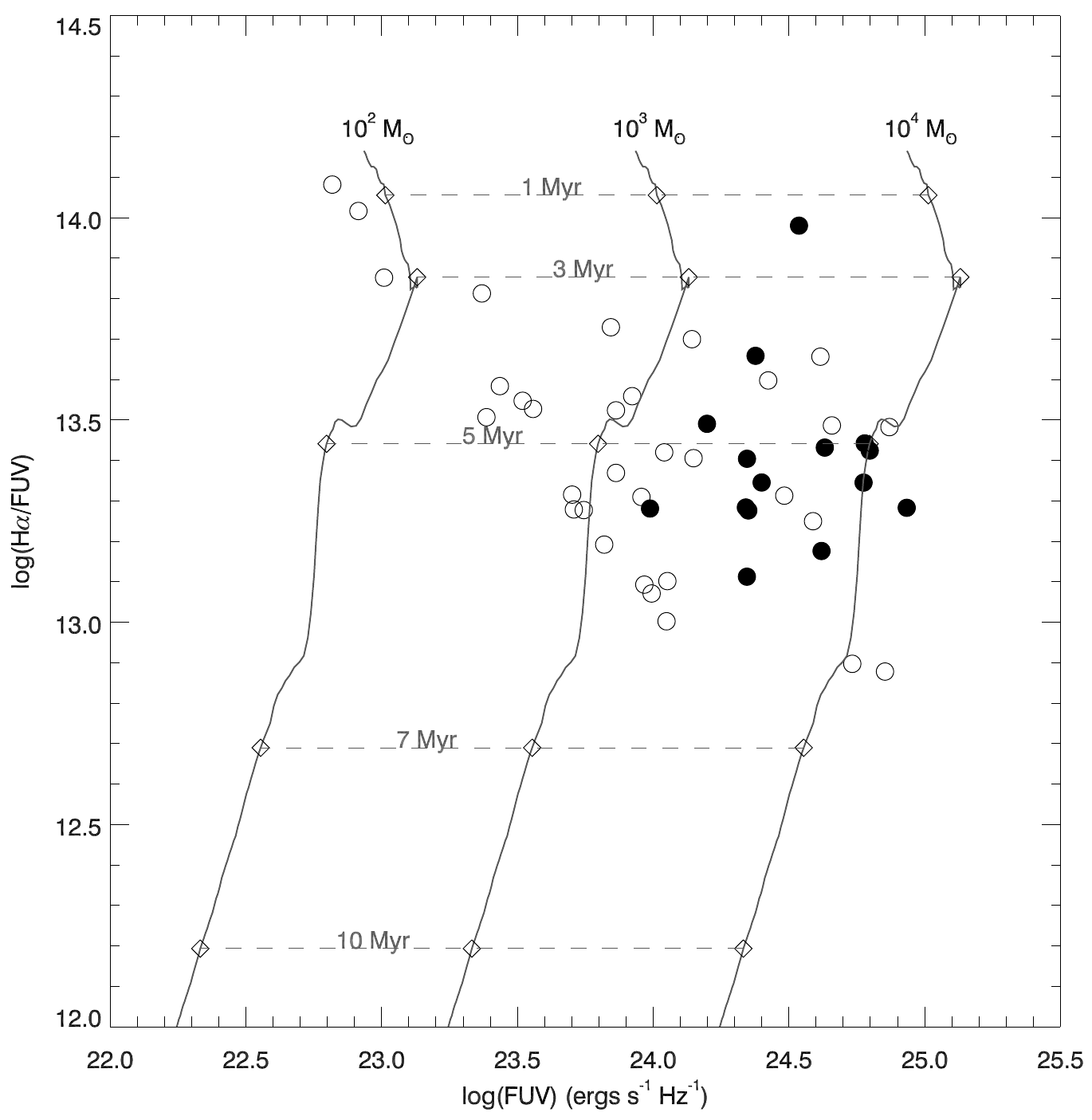} 
     \caption{FUV/\ha\ ratio as a function of FUV luminosity for XUV disc objects in NCG 4625 as measured by \citet{goddard10}. Objects for which an \hii\ region in our sample falls under the area of a UV-detected object are shown as filled circles. Open circles are used in cases where we have instead measured the total \ha\ emission coming
from a region that matches in size the UV-identified objects. Grey lines show theoretical Starburst99 models for three cluster masses (10$^2$, $10^{3}$ and $10^4$ \msun). Dashed lines connect models of constant age.}
   \label{fig:hiiage}
\end{figure}

We  estimated the ages of the \hii\ regions in our sample by looking at  the UV and \ha\ fluxes measured by \citet{goddard10}. In Fig.~\ref{fig:hiiage}  we plot the FUV/\ha\ ratio as a function of the FUV luminosity. We have marked with filled circles those objects for which an \hii\ region matches the position of an identified FUV object. For the remaining cases (open circles) we have measured the total \ha\ emission coming
from a region that matches in size the UV-identified objects.
We include  Starburst99 models calculated for a single burst of star formation at three different cluster masses (10$^2$, $10^{3}$ and $10^4$~\msun). Dashed lines connect models of identical age. 
It can be seen that the cluster ages predicted by the models are  roughly consistent with the ages of the {\sc itera} models at which deviations in the BPT diagram occur. However,  the data are unable to break the age-stochasticity degeneracy,  and thus we cannot differentiate between the two effects. 

Fig.~\ref{fig:hiiage} also provides estimates for the embedded cluster masses, indicating values ranging between $10^3$ and $10^4$ \msun. Such low cluster masses are in the regime that is under the influence of stochastic variations in the upper end of the stellar mass function, as already pointed out by \citet{gildepaz05}.

\subsubsection{Geometry}

In an \hii\ region with a small number of ionising sources, such as those which have a stochastically selected population of massive stars, the spatial distribution of the ionising sources relative to the nebular gas can affect the ionizing properties of the cluster. 
\citet{ercolano07}  showed that geometric effects could produce theoretical \hii\ regions which lie off the main ionising sequence in the BPT diagnostic diagrams, as well as affect the nebular abundance determination, especially at low metallicities. However, no particular model from this paper was found to reproduce our observations.

\subsubsection{Anomalous N/O ratios}

In the outer discs of galaxies the star formation rate is particularly low,  typically  only a few percent of that in the inner discs, moreover spread over a much larger area \citep{gildepaz05,goddard10}. In proposing that small stellar clusters are unable to form massive stars, \citet{kroupa08} suggested that this might explain the edge of \ha\ emission in galactic discs and the lack of massive stars beyond this radius. With a dearth of massive stars in the outer discs,  and therefore of core-collapse supernova explosions, we would expect a reduced enrichment of oxygen relative to nitrogen, since oxygen is produced by the most massive (M~$>$~8~\msun) stars only, while nitrogen is also produced by intermediate-mass stars. Could such an effect be responsible, at least in part, for the anomaly that we found in the BPT diagram?

The left-hand panel of Fig.~\ref{fig:metcheck} shows that the deviant \hii\ regions  in the BPT diagnostic diagram (green dots)  are shifted to the left relative to the sequence formed by bright \hii\ regions also in the 
R$_{23}$ vs. \nii/\oii\ plane. The shift appears to occur at constant  \nii/\oii\ ratio, at variance with what  one would expect if the gas was lacking in oxygen relative to nitrogen, which would move the objects towards the top of the diagram. 
 
\subsection{Reliability of the Strong Line Abundances} \label{sec:metcheck}

We still need to consider the fact that the systematic offsets found in a subset of the outer disc \hii\ regions in NGC~4625 could, in principle, affect the abundances determinations based on strong line indicators. Would these indicators, developed for bright extragalactic ionized nebulae, still be applicable  to much fainter \hii\ regions, where the population  of massive stars is stochastically determined? 

\begin{figure*}
   \centering
   \includegraphics[width = 8.8cm]{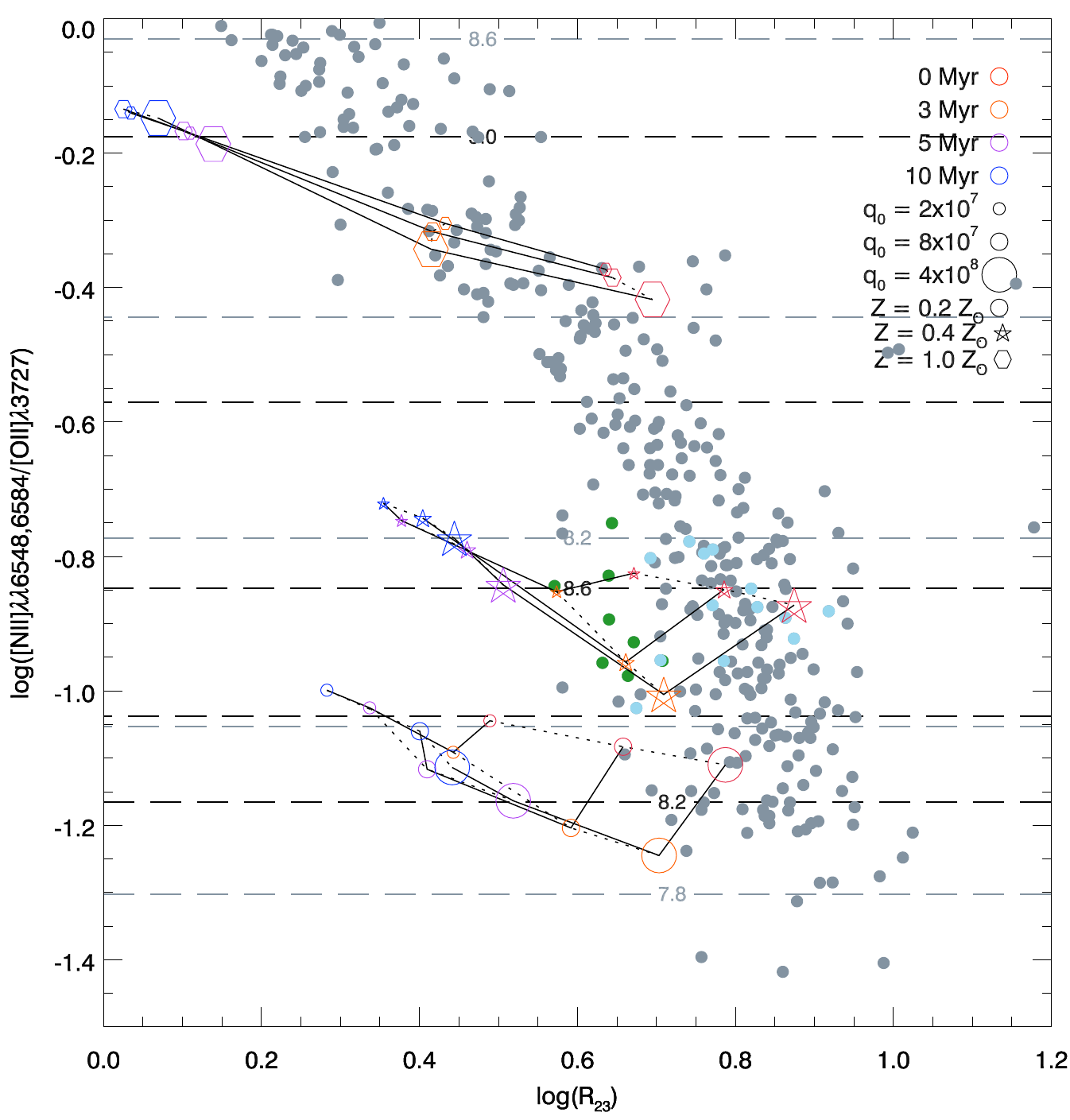} 
   \includegraphics[width = 8.8cm]{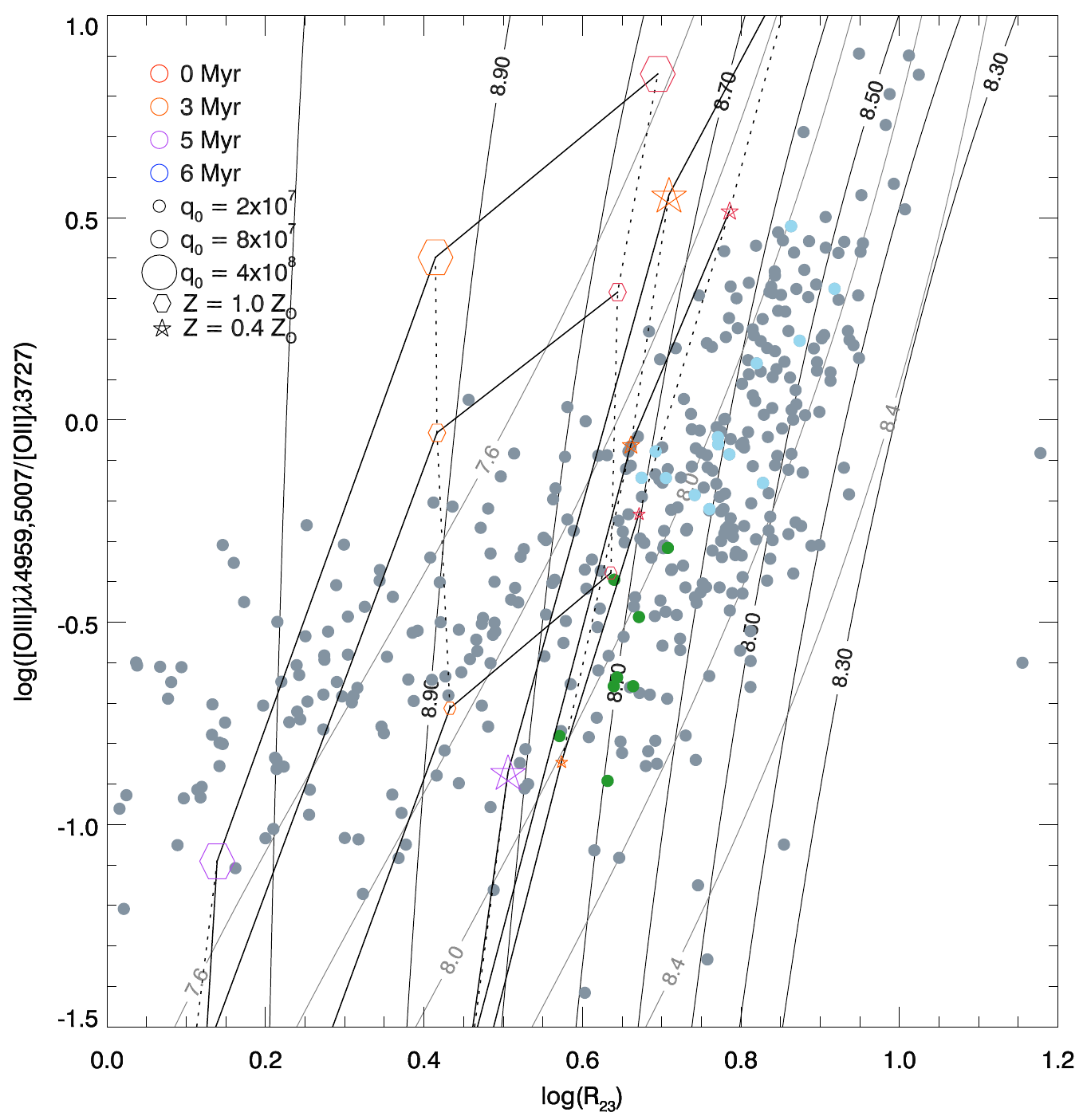} 
   \caption{{\em Left:} log R$_{23}$ vs. log \nii/\oii. We show lines of constant metallicity based on the N2O2 calibrations of \citet[black dashes]{kewley02} and \citet[grey dashes]{bresolin07}. 
{\em Right:} log~R$_{23}$ vs. log~\oiii/\oii. We show lines of constant metallicity, based on the upper (black lines) and lower (grey) branches of the R$_{23}$ calibration of \citet{mcgaugh91}.
In both panels the grey dots show the same sample of \hii\ regions plotted in previous figures. Cyan dots represent \hii\ regions from our NGC~4625 sample for which \oiii/H$\beta > 0.25$, whilst green dots represent those with \oiii/H$\beta < 0.25$. We have also plotted {\sc itera} nebular models using the same symbols as in Fig.~\ref{fig:mod1}. }
   \label{fig:metcheck}
\end{figure*}

In Fig.~\ref{fig:metcheck} we show model predictions and empirical data for R$_{23}$ as a function of  the \nii/\oii\ ratio (left panel) and of  the \nii/\oii\ ratio (right panel). In the 
latter case we plotted lines of constant metallicity, adopting both the upper (black lines) and lower (grey) branch of the theoretical  calibration by \citet{mcgaugh91}.

The left-hand panel of Fig.~\ref{fig:metcheck} gives us two major insights. Firstly, it confirms our earlier finding (\S \ref{sec:hiiage}) that the line emission properties of ageing \hii\ regions appear to replicate those of  the \hii\ regions in our sample which were found to lie off the main ionisation sequence in the BPT diagnostic diagram (green dots). Secondly, it shows how an ageing population of \hii\ regions might affect the determination of the chemical abundances. The {\sc itera} models suggest that as an \hii\ region ages, its R$_{23}$ value rapidly decreases, resulting in a shift almost parallel to the lines of constant metallicity (determined using the N2O2 diagnostic). The 
\nii/\oii\ ratio also changes, simulating a variation in the inferred metallicity.
 Therefore, the models give us an indication of how reliable the metallicity indicators become for an ageing \hii\ region population. 
 A spread of ages of, say, between 3 and 6 Myr, as suggested by Fig.~\ref{fig:hiiage} for NGC~4625,
could result in an uncertainty in the abundance measurement of roughly $\sim0.15$~dex, based on the N2O2 diagnostic.

The right-hand panel of Fig.~\ref{fig:metcheck} helps us to identify how robust the R$_{23}$ indicator may be against such an ageing \hii\ region population. For a metallicity of 0.4 \zsun\  the \hii\ regions appear to evolve almost parallel to the lines of constant metallicity, although there may be some divergence with the lower branch calibration for the oldest models. However, for the 1\,\zsun\ metallicity case the \hii\ region evolution appears to be problematic for the models, since large changes in the measured abundance, of the order $\sim$0.25 dex,
appear to occur, together with discrepancies for models of the same age but lower ionisation parameters. 
This suggests  that for an ageing population of \hii\ regions with a spread of ionisation parameters  the R$_{23}$ indicator could represent an inaccurate measure of the oxygen abundance.

\section{Discussion} \label{sec:interp}

\subsection{Breaks in Abundance Gradients}

In \S \ref{sec:slabunds} we showed that the strong line abundance indicators used in this study gave qualitatively similar results. All showed a discontinuity in the abundance gradient at a galactocentric distance between 0.7 and 1.1 R$_{25}$, and a shallower gradient in the outer disc. 

While the abundance solution provided by the use of the R$_{23}$ indicator remains somewhat dubious, due to the uncertainty in the choice between upper and lower branch calibration, the N2 and N2O2 abundance diagnostics 
are monotonic with abundance and are unlikely to produce false discontinuities or gradient changes. 
In addition, \citet{bresolin09} tested the reliability of N2O2 to measure relative abundances in both the inner and outer disc of M83. 

In both NGC~4625 (this work) and in M83 (\citealt{bresolin09}) we found a discontinuity in the abundance gradient, with  a shallower slope beyond R$_{25}$.
\citet{goddard10} also found similar radial UV and \ha\  profiles in the two galaxies. The FUV profile is smooth over the transition between inner and outer disc, becoming  shallower beyond R$_{25}$. Conversely, the \ha\ profiles show a sharp truncation at R$_{25}$, in correspondence with the edge of the star forming disc.
The origin of these effects lies possibly in the differing star formation properties between inner and outer parts of the discs we have analyzed.
The extremely low  \ha\ emission in the extended discs indicates a very low star formation rate \citep{gildepaz05,goddard10}.
The lack of massive stars results in a smaller chemical enrichment from supernova events, which would eventually lead  to  the oxygen abundance discontinuity observed between the inner and outer discs. G07 have argued that both a low-level, but continuous, star formation activity and an episodic one could be invoked to explain the fact that the oxygen abundances are low, but not extremely so, in the extended discs of spirals (see also \citealt{bresolin09}). \citet{bresolin09} have remarked on the similarity between XUV discs and low-surface brightness galaxies in terms of chemical abundances and enrichment. They also pointed out that gravitational interactions with companion galaxies could generate, via gas stripping, peculiarities in the abundance gradients, such as those observed in the outer disc of M83, which has likely interacted with NGC~5253 around 1~Gyr ago. The situation appears to be similar in the case of NGC~4625, which has  been in interaction with its companion NGC~4618 and, possibly, a fainter galaxy (\citealt{gildepaz05}). Perhaps these interactions play an important role both in the activation of the star formation in the extended discs, and in the generation of  the abundance breaks and/or flattened gradients.

\subsection{H\,{\small II} Region populations in XUV discs}

We found a significant number of \hii\ regions in NGC~4625 that appear to depart from the main ionisation sequence established by studies of bright, inner disc ionized nebulae. A similar behaviour is detected (Fig.~\ref{fig:bpt}) for a small fraction of \hii\ regions in other galaxies.  
The most likely reasons for this anomaly have been identified in stochasticity or the effects of an ageing \hii\ region population. Since the outer disc \hii\ regions contain at most only a few O-type stars,  the source of ionizing flux is subject to stochastic variations \citep{gildepaz05,thilker05,goddard10}. By using measurements from \citet{goddard10}, in combination with published photoionisation models, we also found that an ageing \hii\ region population might equally be responsible for the spread  we observe in the BPT diagnostic diagram. These two effects are indistinguishable, since an evolving \hii\ region loses first the most massive stars, which are also those that are most affected by stochastic variations. 

We examined the possibility that  geometry or a non-standard N/O ratio might explain the emission line properties of the deviant \hii\ regions. However, we concluded that these effects are unlikely to produce the observed trends, and both age and stochastic arguments appear to be more robust. It was never our intent to produce a full description of these effects, but only to highlight the fact that outer disc \hii\ regions are significantly different from their inner disc counterparts, and this difference has not been fully investigated yet. In the future it will be important to
investigate the presence of \hii\ regions in other galaxies with emission line properties similar to what we found in NGC~4625,
and carry out new comparisons to models generated with a stochastically populated IMF.

\section{Conclusions} \label{sec:conc}

In this study we have analysed the spectra of 34 \hii\ regions at a range of galactocentric radii in the XUV disc galaxy NGC~4625. Below we summarise our main findings:

\begin{description}
	\item[1.] Strong line abundance indicators consistently show a shallower abundance gradient in the outer disc compared to the inner disc. This is accompanied by a discontinuity in the abundance gradient that occurs in proximity of  the optical edge of the galaxy. These trends are very similar to those seen in  M83.
	\item[2.] The three \te-based abundances in the outer disc predict a metallicity of \oh\ $\simeq 7.96\,\pm\,^{0.22}_{0.31}$ ($\sim 0.2$ \zsun) in the outer disc.
	\item[3.] We identified a subset of outer disc \hii\ regions which do not conform to the ionisation sequence defined by the bright inner disc \hii\ regions in the \oiii/\hbeta\ \vs\ \nii/\halpha\ diagnostic diagram. These objects do not appear to display any unique properties in terms of position within the XUV disc or ionising flux.
	\item[4.] Stochastic variations in the high-mass end of the embedded stellar population or an ageing \hii\ region population seem to be the most likely effects capable of producing the observed  emission line properties. However, it has not been possible to distinguish between these two effects.
	\item[5.] The strong line abundance indicators used in this paper appear to be less reliable if the \hii\ region population is ageing, even over the short 10 Myr timescale of the ionising stars. This effect could be as much as $\sim$ 0.15 dex for the N2O2 indicator, and even more for R$_{23}$.
	
\end{description}

We hope future studies of outer disc \hii\ regions will be able to further probe this regime of stochastically determined ionisation sources. 
It is only through  the continued study of the abundance gradients in a variety of XUV discs that we may hope to better understand the connection between star formation and chemical enrichment of the interstellar medium in these low gas density environments.

\bigskip
\bigskip
\noindent
FB gratefully acknowledges the support from the National Science Foundation grant AST-0707911.

\bibliography{master}{}
\bibliographystyle{mn2e}

\bsp

\label{lastpage}

\end{document}